\documentclass[aip, jcp, notitlepage, floatfix, reprint,
  citeautoscript, twocolumn]{revtex4-1}

\usepackage{amsmath}
\usepackage{amsfonts}
\usepackage{graphicx}
\usepackage{dcolumn}
\usepackage[table]{xcolor}
\usepackage[version=3]{mhchem}
\usepackage{transparent}
\usepackage{gensymb}
\usepackage{tabularx}
\usepackage{etoolbox}
\usepackage[hidelinks]{hyperref}
\usepackage{bm}
\usepackage{relsize}
\usepackage[normalem]{ulem}
\usepackage{cases}

\def\maxfloatwidth{%
  \ifdim\columnwidth>246.0pt
  300.0pt  \else
  \columnwidth
  \fi
}

\newcommand{\tbf}[1]{\textbf{#1}}

\newcommand{\trm}[1]{\textrm{#1}}
\newcommand{\mrm}[1]{\mathrm{#1}}
\newcommand{\mbf}[1]{\mathbf{#1}}
\newcommand{\tcr}[1]{\textcolor{black}{#1}}

\newcommand{\<}{\langle}
\renewcommand{\>}{\rangle}
\newcommand{\beq}{\begin{equation}}
\newcommand{\eeq}{\end{equation}}

\definecolor{bgpeach}{rgb}{1.000,0.925,0.850}
\definecolor{fggray}{rgb}{0.384,0.435,0.471}

\begin{document}


\title{Assessing long-range contributions to the charge asymmetry of
  ion adsorption at the air-water interface}

\author{Stephen J. Cox}
\affiliation{Department of Chemistry, University of Cambridge,
  Lensfield Road, Cambridge CB2 1EW, United Kingdom}

\author{Dayton G. Thorpe}
\affiliation{Chemical Sciences Division, Lawrence Berkeley National
  Laboratory, Berkeley, CA 94720, United States.}
\affiliation{Department of Physics, University of California,
  Berkeley, CA 94720, United States.}

\author{Patrick R. Shaffer}
\affiliation{Department of Chemistry, University of California,
  Berkeley, CA 94720, United States.}

\author{Phillip L. Geissler}
\affiliation{Chemical Sciences Division, Lawrence Berkeley National
  Laboratory, Berkeley, CA 94720, United States.}
\affiliation{Department of Chemistry, University of California,
  Berkeley, CA 94720, United States.}

\date{\today}

\begin{abstract}
Anions generally associate more favorably with the air-water interface
than cations. In addition to solute size and polarizability, the
intrinsic structure of the unperturbed interface has been discussed as
an important contributor to this bias. Here we assess quantitatively
the role that intrinsic charge asymmetry of water's surface plays in
ion adsorption, using computer simulations to compare model solutes of
various size and charge. In doing so, we also evaluate the degree to
which linear response theory for solvent polarization is a reasonable
approach for comparing the thermodynamics of bulk and interfacial ion
solvation. Consistent with previous works on bulk ion solvation, we
find that the average electrostatic potential at the center of a
neutral, sub-nanometer solute at the air-water interface depends
sensitively on its radius, and that this potential changes quite
nonlinearly as the solute's charge is introduced.  The nonlinear
response closely resembles that of the bulk. As a result, the net
nonlinearity of ion adsorption is weaker than in bulk, but still
substantial, comparable to the apparent magnitude of macroscopically
nonlocal contributions from the undisturbed interface. For the
simple-point-charge model of water we study, these results argue
distinctly against rationalizing ion adsorption in terms of surface
potentials inherent to molecular structure of the liquid's boundary.
\end{abstract}

\maketitle

Counter to expectations from conventional theories of solvation, there
is a large body of both computational and experimental evidence
indicating that small ions can adsorb to the air-water
interface.\cite{jungwirth2006specific,netz2012progress,OttenSaykally2012sjc,Mucha2005unified,petersen2005enhanced,piatkowski2014extreme,verreault2012conventional,liu2004vibrational,baer2011toward}
Implications across the biological, atmospheric and physical sciences
have inspired efforts to understand the microscopic driving forces for
ions associating with hydrophobic interfaces in
general.\cite{ben2016interfacial,Noah-VanhouckeGeissler2009sjc,arslanargin2012free,baer2014toward,beck2013influence,dos2013surface,levin2009polarizable,levin2009ions,mccaffrey2017mechanism,ou2013spherical,ou2013temperature,caleman2011atomistic}
A particular emphasis has been placed on understanding ion
specificity, i.e., why some ions exhibit strong interfacial affinity
while others do not.  Empirical trends indicate that ion size and
polarizability are important factors, as could be anticipated from
conventional theory.  More surprisingly, the {\em sign} of a solute's
charge can effect a significant bias, with anions tending to adsorb
more favorably than cations.

Here we examine the microscopic origin of this charge asymmetry in
interfacial ion adsorption. We specifically assess whether the
thermodynamic preference can be simply and generally understood in
terms of long-range biases that are intrinsic to an aqueous system
surrounded by vapor. By ``long-range'' and ``nonlocal'' we refer to
macroscopically large scales, i.e., collective forces that are felt at
arbitrarily long distance.  Such a macroscopically long-range bias is
expected from the air-water interface due to its average polarization,
and by some measures the bias is quite strong.  By contrast, ``local''
contributions comprise the entire influence of a solute's microscopic
environment, including electrostatic forces from molecules that are
many solvation shells away -- any influence that decays over a
sub-macroscopic length scale.

The importance of macroscopically nonlocal contributions has been
discussed extensively in the context of ion solvation in bulk liquid
water, which we review in Sec.~\ref{sec:BulkReview} as a backdrop for
interfacial solvation. The notion that such contributions strongly
influence charge asymmetry of solvation at the air-water interface has
informed theoretical approaches and inspired criticism of widely used
force fields for molecular
simulation.\cite{baer2012electrochemical,levin2014ions} A full
understanding of their role in interfacial adsorption, however, is
lacking.

In the course of this study, we will also evaluate the suitability of
dielectric continuum theory (DCT) to describe the adsorption process.
DCT has provided an essential conceptual framework for rationalizing
water's response to electrostatic perturbations. But a more precise
understanding of its applicability is needed, particularly for the
construction of more elaborate models (e.g., with heterogeneous
polarizability near
interfaces\cite{loche2018breakdown,loche2020universal,schlaich2016water})
and for the application of DCT to evermore complex (e.g., nanoconfined
\cite{fumagalli2018anomalously,bocquet2020nanofluidics}) environments.

\section{Charge asymmetry in bulk liquid water}
\label{sec:BulkReview}

Our study of {\em interfacial} charge asymmetry is strongly informed
by previous work on the solvation of ions in {\em bulk} liquid water.
In this section we review important perspectives and conclusions from
that body of work, as a backdrop for new results concerning ions at
the air-water interface.

\subsection{Distant interfaces and the neutral cavity potential}

A difference in adsorption behaviors of anions and cations is
foreshadowed by the fact that ion solvation in models of bulk liquid
water is also substantially charge asymmetric. Born's classic model
for the charging of a solute captures the basic scale of solvation
free energies, as well as their rough dependence on a solute's
size.\cite{latimer1939free} We will characterize the size of a solute
by its radius $R$ of volume exclusion, the closest distance that a
water molecule's oxygen atom can approach without incurring a large
energetic penalty. Contrary to Born's result, computer simulations
indicate that the sign of the charge of small ions can significantly
influence their charging free energy $F_{\rm chg}(q,R)$ i.e., the work
involved in reversibly introducing the solute's charge
$q$.\cite{duignan2017real,duignan2017electrostatic,remsing2016role,bardhan2012affine,HummerGarcia1996sjc,rajamani2004size,lynden1997hydrophobic,cox2018interfacial,ashbaugh2008single,remsing2019influence}
This dependence is most easily scrutinized for simple point charge
(SPC) models of molecular interactions, where an ion's charge can be
varied independently of its other properties.  In SPC/E
water,\cite{BerendsenStraatsma1987sjc} for instance, charging a solute
roughly the size of fluoride ($R_{\rm F} \approx 0.317$\,nm) has an
asymmetry, $F_{\rm chg}(e,R_{\rm F}) - F_{\rm chg}(-e,R_{\rm F})
\approx 16$\,kcal/mol, almost 30 times larger than thermal energy
$k_{\rm B}T$. Here, $e$ is the magnitude of an electron's charge.

The ultimate origin of charge asymmetry in liquid water is of course
the inequivalent distribution of positive and negative charge in a
water molecule itself. On average, the spatial distribution of
positive and negative charge is uniform in the bulk liquid, but any
breaking of translational symmetry will manifest the distinct
statistics of their fluctuating arrangements.  A neutral, solute-sized
cavity in water, for example, experiences an immediate environment in
which solvent molecules have a nonvanishing and spatially varying net
orientation. The internal charge distributions of these oriented
solvent molecules generate a nonzero electric potential at the center
of the cavity, whose sign and magnitude are not simple to anticipate.
By our characterization, this electrostatic bias is {\em local} in
origin -- the total contribution of molecules beyond a distance $r$
from the cavity decays to zero as $r$ increases.

The inequivalent spatial distribution of positive and negative charge
in water can generate spatially {\em nonlocal} biases as well, effects
that extend over arbitrarily large distances.  Any point in the bulk
liquid is macroscopically removed from the physical boundaries of the
liquid phase (e.g., interfaces with a coexisting vapor phase), but
those distant boundaries may nonetheless impact the thermodynamics of
bulk ion solvation.  This expectation stems from a textbook result of
electrostatics: an infinitely extended (or completely enclosing)
dipolar surface, with polarization pointing along the surface normal,
generates a discontinuity in electric potential. This voltage offset
does not decay with distance from the interface, and thus meets our
criterion for macroscopic nonlocality.  A two-dimensional manifold of
polarization density is certainly a crude caricature of a liquid-vapor
interface, but for a polar solvent whose orientational symmetry is
broken at its boundaries, a similarly long-range potential from the
interface is expected to bias the solvation of charged solutes, even
macroscopically deep inside the liquid phase.

\begin{figure}[!tb]
  \centering
  \includegraphics[width=7.68cm]{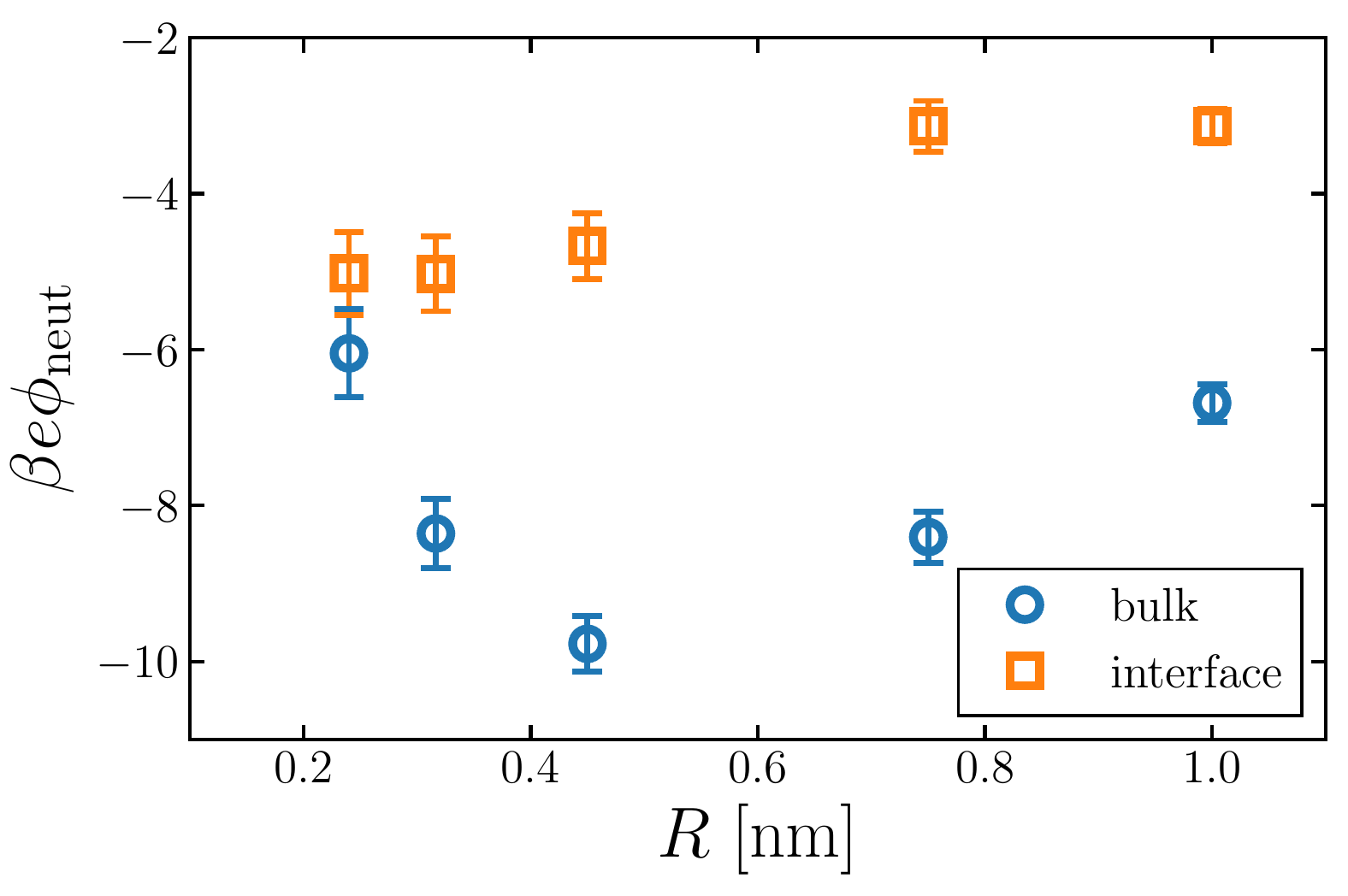}
  \caption{ The average electric potential $\phi_{\rm neut}$ at the
    center of a neutral cavity varies considerably with the cavity's
    radius $R$. Moreover, this dependence differs for the solute at
    $z=z_{\rm liq}$ (``bulk'') and $z=z_{\rm int}$
    (``interface''). The error bars indicate 95\% confidence
    intervals.}
  \label{fig:phi_neut}
\end{figure}

The average electric potential $\phi_{\rm neut}$ at the center of a
neutral cavity, which we call the ``neutral cavity potential'', sums
these local and extremely nonlocal contributions. The former depends
on the cavity's size (or more generally on the geometry of the solute
represented by the cavity). The latter, interfacial contribution
should, by contrast, be insensitive to such microscopic details, since
the distant surface is unperturbed by the solute.  The net
electrostatic bias from these two sources can be straightforwardly
calculated in computer simulations, not only for SPC models but also
with \emph{ab initio}
approaches.\cite{beck2013influence,remsing2014role,duignan2017real,duignan2017electrostatic}
Fig.~\ref{fig:phi_neut} shows $\phi_{\rm neut}(R)$ for cavities in
bulk liquid SPC/E water (properly referenced to vapor following
Ref.~\citenum{cox2018interfacial}).  Negative potentials of a few
hundred mV, varying by nearly a factor of two as $R$ grows from
0.2\,nm to 1\,nm, echo results of previous
studies.\cite{ashbaugh2000convergence} Distinguishing quantitatively
between local and nonlocal contributions to $\phi_{\rm neut}$,
however, is surprisingly confounding, even for the exceedingly strict
definition of nonlocality considered here.

One strategy to remove local contributions from $\phi_{\rm neut}$ is
to consider the limit $R=0$. In this extreme case the probe -- in
effect a neutral, non-volume excluding solute -- does not break
translational symmetry and induces no structural response.  Given the
lack of local structure, the presumably nonlocal quantity $\phi_{\rm
  neut}(0) = \phi_{\rm surf}$ is often called the ``surface
potential''.  Lacking volume exclusion, however, this probe explores
the liquid phase uniformly, including even the interior of solvent
molecules where electrostatic potentials can be very large.  A
disturbing ambiguity results: The value of $\phi_{\rm surf}$ can be
sensitive to modifications of a solvent model that have no impact on
the solvation of any volume-excluding
solute. Refs.~\citenum{remsing2019influence}
and~\citenum{remsing2014role} illustrate this issue vividly,
constructing `smeared shell' variants of SPC models with identical
solvation properties but very different values of $\phi_{\rm surf}$.
This variation in surface potential corresponds to differences in the
so-called Bethe potential, which is discussed further in the
Supporting Information (SI).

A related, and somewhat more molecular, approach to isolating the
electrostatic bias from a distant phase boundary is to sum
contributions to $\phi_{\rm neut}$ only from molecules that reside in
the interfacial region. For a macroscopic droplet of liquid water, one
could classify each molecule in a given configuration as either
interfacial or bulk based on its position relative to the
interface. The restricted sum
\begin{equation}
  \phi_{\rm d} = \left\langle
  \sum_{j\in {\rm interface}}^N \sum_\alpha \frac{q_\alpha}{|{\bf r}_{j\alpha}|}
  \right\rangle
  \label{equ:phid}
\end{equation}
could then be considered as a macroscopically long-ranged,
surface-specific component of $\phi_{\rm surf}$ that is appropriately
insensitive to a solvent molecule's internal structure. Here ${\bf
  r}_{j\alpha}$ denotes the position of site $\alpha$ in molecule $j$,
whose charge is $q_\alpha$, relative to the center of the droplet.
$\phi_{\rm d}$ depends significantly, however, on the way molecules
are notionally divided between surface and bulk.  This dependence,
which has been demonstrated previously,
\cite{aaqvist1998analysis,kastenholz2006computation} we calculate
explicitly and generally in the SI.  Written in the form
$$
\phi_{\rm d} = -4\pi \int_{z_{\rm liq}}^{z_{\rm vap}}\!dz\,P(z)
$$
where $P(z)$ is the solvent dipole density at a displacement $z$ from
the interface, it reveals $\phi_{\rm d}$ as the well-known ``dipole
component'' of the surface
potential.\cite{remsing2014role,remsing2019influence,duignan2017real,duignan2017electrostatic,HunenbergerAndReifBook,doyle2019importance,horvath2013vapor}
Here, $z_{\rm liq}$ and $z_{\rm vap}$ indicate points within the bulk
liquid and bulk vapor, respectively.

For SPC/E water, a surface/bulk classification in Eq.~\ref{equ:phid}
based on the position of a water molecule's center of charge gives a
value $\phi_{\rm d}^{\rm center}=-40$\,mV that differs from an oxygen
atom-based classification, $\phi_{\rm d}^{\rm O}=240$\,mV, even in
sign.\footnote{We define a molecule's center of charge according to
  the charged sites that specify a particular SPC model. In the case
  of SPC/E water, this center is displaced from the oxygen atom by
  approximately 0.029\,nm along the molecular dipole.} Because water
molecules are not point particles, there is no unique way to define an
interfacial population, and as a result no unique value of $\phi_{\rm
  d}$, though attempts have been made to define an optimal
choice.\cite{kastenholz2006computation} And because molecules near the
liquid's boundary are not strongly oriented on average, the range of
plausible values for $\phi_{\rm d}$ is as large as their mean.

The ambiguities plaguing interpretations of $\phi_{\rm surf}$ and
$\phi_{\rm d}$ are one and the same. Indeed, if we consider an
interfacial population of charged sites rather than intact molecules,
then $\phi_{\rm surf}$ and $\phi_{\rm d}$ become equal. (When defining
an interface of intact molecules, $\phi_{\rm surf}$ and $\phi_{\rm d}$
differ by the so-called Bethe potential, whose analogous ambiguity is
described in SI.) $\phi_{\rm neut}$ has been characterized as a
two-interface
quantity,\cite{remsing2014role,harder2008origin,beck2013influence,arslanargin2012free,doyle2019importance,horvath2013vapor}
combining the bias $\phi_{\rm d}$ from the distant solvent-vapor
interface together with the remaining ``cavity'' bias $\phi_{\rm
  c}=\phi_{\rm neut}-\phi_{\rm d}$ from the local solute-solvent
interface.  From the perspective we have described, these two
interfaces are not truly separable, even if a macroscopic amount of
isotropic bulk liquid intervenes between them -- they must be defined
consistently, and the manner of definition substantially influences
the change in electrostatic potential at each interface. This is not
to say that such a decomposition cannot be useful. Indeed, for
computationally demanding \emph{ab initio} approaches it can be
convenient to consider local and nonlocal contributions to $\phi_{\rm
  neut}$ such that, in a first step, $\phi_{\rm c}$ can be obtained
from relatively small simulations of the bulk under periodic boundary
conditions. The effects of $\phi_{\rm d}$ can then be accounted for in
a subsequent step involving simulations of the neat air-water
interface. Such an approach was used to good effect in
Ref.~\citenum{duignan2017real} to calculate the solvation free energy
of LiF. Nonetheless, this still amounts to an arbitrary choice of
dividing
surface,\cite{duignan2017real,remsing2014role,remsing2019influence}
making it challenging to assign a physical interpretation to
$\phi_{\rm d}$ and $\phi_{\rm c}$ individually. Different, and equally
plausible, ways of partitioning molecules can give different
impressions of the two interfaces. Only the sum $\phi_{\rm neut} =
\phi_{\rm c} + \phi_{\rm d}$ is unambiguous.

Establishing an absolute electrostatic bias on the bulk liquid
environment due to a distant interface is thus highly problematic for
water. A direct scrutiny of this nonlocal contribution, based on the
fundamentally ambiguous potential $\phi_{\rm d}$, is untenable.
Instead, we assess the relative importance of local and nonlocal
biases by comparing the solvation properties of different ions.  Local
contributions can depend sensitively on features like solute size $R$
and charge $q$, while macroscopically nonlocal contributions cannot.
Long-range influence of the interface might therefore be clarified by
dependence of the neutral cavity potential on $R$. In particular,
dominance by the distant liquid-vapor interface would imply weak
variation of $\phi_{\rm neut}$ with solute size, which influences only
microscopically local structure. The solute size-dependence shown in
Fig.~\ref{fig:phi_neut} does not support such a dominance. Growing the
cavity from $R=0.24$\,nm to $0.5$\,nm lowers $\phi_{\rm neut}$ by
roughly 100\,mV, followed by an increasing trend for larger
cavities. As emphasized in Refs.~\citenum{duignan2017real}
and~\citenum{remsing2014role}, the role of local charge asymmetry is
far from negligible over this range of solute size.

It is tempting to expect the large-$R$ behavior of $\phi_{\rm neut}$
to reveal a strictly interfacial component, since local forces
attenuate in magnitude when solvent molecules cannot approach the
probe position closely. As others have
noted,\cite{remsing2014role,ashbaugh2000convergence} however, neutral
cavities larger than $R=1$\,nm induce a solvent environment with the
basic character of the air-water
interface.\cite{chandler2005interfaces} In the limit of large $R$,
drying at the solute-solvent interface will generate a cavity
potential that cancels the oppositely oriented distant interface with
the vapor phase, yielding $\phi_{\rm neut}\approx 0$.\footnote{While
  the vapor phase is very dilute at ambient temperature, its nonzero
  density does yield an average potential different from the vacuum
  environment of a volume-excluding cavity. Here we neglect this small
  distinction.}  This asymptotic cancellation should begin for
nanoscale cavities, though effects of local interface curvature may
cause $\phi_{\rm neut}$ to decay slowly towards zero.  Judging from
our results, there is no intermediate plateau value of $\phi_{\rm
  neut}$ that could reasonably be assigned to a single liquid-vapor
interface.

\subsection{Solvation thermodynamics and the asymmetry potential}

The difficulty of uniquely identifying a surface dipole component of
$\phi_{\rm neut}$ notwithstanding, the relevance of such neutral probe
quantities for ion solvation thermodynamics has also been thoroughly
examined.\cite{duignan2017real,duignan2017electrostatic,remsing2016role,HummerGarcia1996sjc,beck2013influence,ben2016interfacial,rajamani2004size,shi2013length,doyle2019importance,remsing2019influence,pollard2016toward,asthagiri2003absolute,horvath2013vapor,pratt1992contact}
As an essential thermodynamic measure of solvation, we examine the
free energy change $F_{\rm solv}(q,R)$ when a solute ion is removed
from dilute vapor and added to the liquid phase.  This change could be
evaluated along any reversible path that transfers the solute between
phases, and different paths can highlight different aspects of solvent
response. For studying charge asymmetry, a particularly appealing path
first creates a neutral, solute-sized cavity in the liquid, with
reversible work $F_{\rm cav}(R)$. The second step, whose free energy
change $F_{\rm chg}(q,R)$ was discussed above, introduces the solute's
charge.\cite{netz2012progress} The charge asymmetry of interest
compares solvating a cation and anion of the same size; since $F_{\rm
  cav}$ is insensitive to the solute's charge, its contribution to
$F_{\rm solv}=F_{\rm cav}+F_{\rm chg}$ cancels in the difference
\begin{eqnarray}
F_{\rm solv}(q,R)-F_{\rm solv}(-q,R) &=& F_{\rm
  chg}(q,R)- F_{\rm chg}(-q,R)
\\
&\equiv& 2 q \psi(q,R) \label{eqn:psi}
\end{eqnarray}
Eq.~\ref{eqn:psi} defines an {\em asymmetry potential} $\psi$, an
analogue of $\phi_{\rm neut}$ that accounts for solvent response.

The connection between $\psi(q,R)$ and $\phi_{\rm neut}$ can be made
precise through a cumulant expansion of $F_{\rm chg}$ in powers of
$q$,\cite{kubo1962generalized,HummerGarcia1996sjc,beck2013influence,ben2016interfacial,rajamani2004size}
\begin{equation}
  \label{eqn:Fchg}
  F_{\rm chg}(q,R) =
  q\langle\phi_{\rm solv}\rangle_{0} -
  \frac{\beta q^2}{2}\langle(\delta\phi_{\rm solv})^2\rangle_{0}
  + \mathcal{O}(q^3),
\end{equation}
where $\langle\cdots\rangle_{0}$ denotes a canonical average in the
presence of a neutral solute-sized cavity, $\phi_{\rm solv}$ is the
fluctuating electric potential at the center of the cavity due to the
surrounding solvent (so that $\phi_{\rm neut}=\langle\phi_{\rm
  solv}\rangle_{0}$), and $\delta\phi_{\rm solv} = \phi_{\rm solv} -
\phi_{\rm neut}$.  The $\mathcal{O}(q^2)$ term in Eq.~\ref{eqn:Fchg}
describes linear response of the solvent potential $\phi_{\rm solv}$
to the solute's charging.  This response, which could be captured by a
Gaussian field theory \`a la DCT, is charge symmetric by construction.
The asymmetry potential $\psi(q,R) = \phi_{\rm neut}(R) +
\mathcal{O}(q^2)$ is therefore equivalent to $\phi_{\rm neut}$ within
linear response.

\begin{figure}[!tb]
  \centering
  \includegraphics[width=7.68cm]{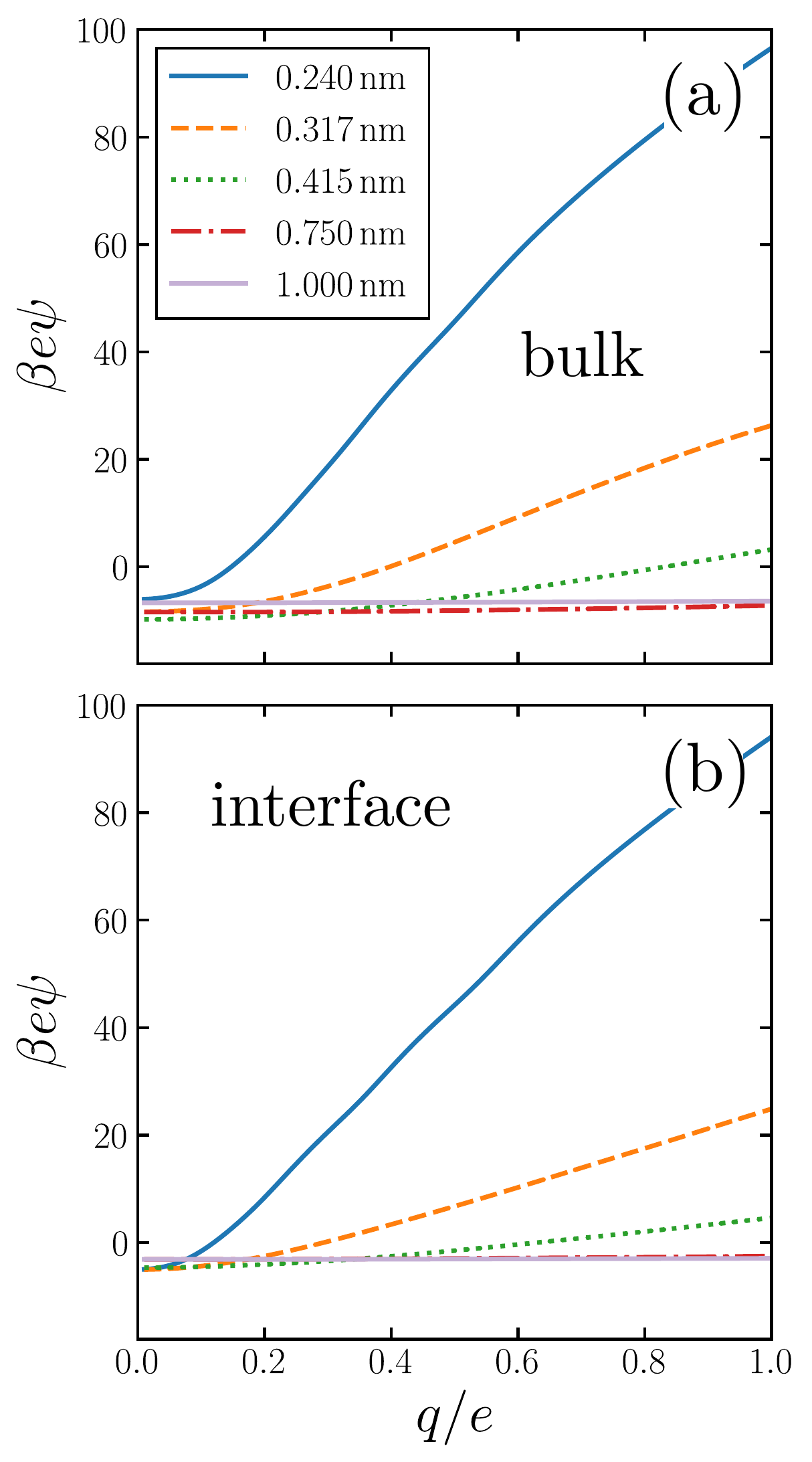}
  \caption{Ion solvation in water is both asymmetric and non-linear,
    as quantified by the asymmetry potential $\psi(q,R;z)$. Results
    are shown for solutes (a) in the bulk liquid, and (b) near the
    air-water interface, spanning ranges of charge $0<q\le e$ and
    solute size $0.24\le R\le 1.0$\,nm (see legend). Both in the bulk
    and at the interface, $\psi<0$ for small $q$, indicating that
    weakly charged cations are more favorably solvated than
    anions. For the smaller solutes, $\psi$ increases with $q$, a
    signature of non-linear response. Anions consequently become more
    favorably solvated at large $q$. For the larger solutes ($R =
    0.75$\,nm and $R=1.0$\,nm) the solvent response is approximately
    linear, as reflected by the weak dependence of $\psi$ on $q$.}
  \label{fig:psis}
\end{figure}

Previous work has demonstrated that water's response to charging
sub-nanometer cavities is significantly
nonlinear.\cite{remsing2016role,bardhan2012affine,HummerGarcia1996sjc,lynden1997hydrophobic,cox2018interfacial,shi2013length,hirata1988viewing,grossfield2005dependence,duignan2017electrostatic,loche2018breakdown}
In $\psi(q,R)$ the breakdown of linear dielectric behavior is
evidenced by deviations away from the limiting value
$\psi(0,R)=\phi_{\rm neut}(R)$.  Fig.~\ref{fig:psis}a shows our
numerical results for the asymmetry potential as a function of $q$ for
solutes in bulk liquid SPC/E water. For large solutes ($R\gtrsim
0.5$\,nm), the variation of $\psi$ is modest as $q$ increases from 0
to $e$.  For smaller cavities, linear response theory fails
dramatically, in that charge asymmetry changes many-fold as the solute
is charged.  In the case of a fluoride-sized solute, the asymmetry at
full charge ($e \psi(e,R_{\rm F})\approx 26 k_{\rm B}T$) is
qualitatively different than in linear response ($e \phi_{\rm
  neut}(R_{\rm F})\approx -8 k_{\rm B}T$). For SPC models of bulk
liquid water, the ultimate electrostatic bias in solvating cations and
anions of this size clearly cannot be attributed to the innate
environment of a neutral cavity, much less to the structure of a
distant interface. \emph{Ab initio} molecular dynamics studies have
reached a similar conclusion.\cite{duignan2017electrostatic}

SPC simulations of bulk liquid water indicate that the nonlinearity of
solvent response to solute charging has a step-like
character:\cite{HummerGarcia1996sjc,lynden1997hydrophobic,bardhan2012affine}
For one range of solute charge ($q<q_{\rm c}$), the susceptibility
$d\<\phi_{\rm solv} \>_q/dq$ is approximately constant. In the
remaining range ($q \geq q_{\rm c}$), $d\<\phi_{\rm solv} \>_q/dq$ is
also nearly constant, but with a different value. Piecewise linear
response (PLR) models inspired by this observation give a broadly
reasonable description of bulk solvation thermodynamics throughout the
entire range $-e<q< + e$. In our discussion of ion adsorption below,
we will assess the suitability of a PLR model for interfacial
solvation as well.

\section{Charge asymmetry in ion adsorption}

In bulk liquid water, an electric potential from its bounding
interfaces cannot be unambiguously identified. Even the sign of the
bias generated by a liquid-vapor interface is unclear.  Moreover, the
nonlinear local response to solute charging can exert a bias on ion
solvation that significantly outweighs the charge asymmetry due to
distant interfaces.

Solvation within the interfacial environment is hardly less complex,
juxtaposing the fluctuating intermolecular arrangements of bulk water
together with broken symmetry and the microscopic shape variations of
a soft boundary. It is thus unlikely that complications described in
Sec.~\ref{sec:BulkReview} for bulk liquid are much eased in the
interfacial scenario. We should not expect, for example, that the
neutral cavity potential for a solute positioned near the interface
will be dominated by a simple nonlocal contribution. Nor should we
expect the accuracy of linear response approximations to be greatly
improved, such that $\phi_{\rm neut}$ is predictive of charge
asymmetric solvation.

The adsorption of an ion to the interface, however, concerns the {\em
  difference} in solvation properties of bulk and interfacial
environments.  To the extent that nonlinear response and local
structuring at the interface are similar to those in bulk liquid,
their effects may cancel, or at least significantly offset, in the
thermodynamics of adsorption. Our main results concern this
possibility of cancellation, which would justify regarding
macroscopically nonlocal contributions to $\phi_{\rm neut}$ as the
basic origin of charge asymmetry in ion adsorption.

We begin by establishing that biases on solvation at the interface are
complicated in ways that qualitatively resemble biases in bulk.  As
before, we consider solutes with a range of sizes and charges, now
positioned at the liquid's boundary (illustrated in Fig. 3a). The free
energies and potentials defined in Sec.~\ref{sec:BulkReview} for bulk
solution now acquire dependence on the Cartesian coordinate $z$ that
points perpendicular to the mean surface. SI shows the detailed
location $z_{\rm int}$ we designate as adsorbed for each ion. In all
cases $z_{\rm int}$ lies near the Gibbs dividing surface, where the
solvent density falls to half its bulk value. The larger solutes
occupy considerable volume, so that the solvent density profile in our
finite simulation cell changes noticeably with their height $z$. A
precise interfacial solute location is therefore difficult to
justify. When neutral and located near $z_{\rm int}$, however, these
nanometer-size solutes tend to deform the instantaneous phase
boundary,\cite{WillardChandler2010sjc,VaikuntanathanGeissler2014sjc,VaikuntanathanGeissler2016sjc}
just as they induce local drying in bulk
solution.\cite{chandler2005interfaces} This response essentially fixes
their location relative to the instantaneous interface, so that their
solvation properties should be fairly insensitive to the choice of
$z_{\rm int}$.

The neutral cavity potential for interfacial solutes is shown in
Fig.~\ref{fig:phi_neut}. As was observed for the bulk liquid,
$\phi_{\rm neut}$ is consistently negative over the range $R=0.24$\,nm
to $R=1$\,nm but varies significantly with solute size. In this case
the potential increases nearly monotonically with $R$, though the
values of $\phi_{\rm neut}(0.75\,{\rm nm})$ and $\phi_{\rm
  neut}(1\,{\rm nm})$ are statistically indistinguishable within our
sampling. Just as for bulk liquid, we expect $\phi_{\rm neut}$ to
vanish in the limit $R\rightarrow\infty$.  Here, drying at the surface
of very large solutes effects a distortion of the liquid-vapor
interface that places the probe (located at the cavity's center)
distinctly in the vapor phase. Judging from our results, the
asymptotic approach to this limit is quite slow for interfacial
solutes. Nonetheless, $\phi_{\rm neut}$ changes by nearly 40\% over
the range of $R$ considered, emphasizing the importance of local,
solute-dependent contributions. As concluded for the bulk solvent,
macroscopically nonlocal potentials arising from orientational
structure of the air-water interface do not dominate the charge
asymmetry experienced by neutral solutes at $z_{\rm int}$.

The response to charging a solute at the air-water interface is
strongly nonlinear, to a degree comparable with bulk response.  A
similarly important role of nonlinear response at interfaces has been
reported
previously.\cite{loche2018breakdown,Noah-VanhouckeGeissler2009sjc} The
resulting $q$-dependent charge asymmetry closely resembles bulk
behavior, as quantified by the asymmetry potential $\psi(q,R;z)$,
whose dependence on solute position we now make explicit.  Fig.~2b
shows simulation results for $\psi(q,R;z_{\rm int})$ for SPC/E
water. On the scale that $\psi$ changes as $q$ increases from 0 to
$e$, the charging response in bulk liquid and at the interface are
nearly indistinguishable by eye. This close similarity suggests that
the predominant source of nonlinearity lies in aspects of local
response which are not so different in the two environments.

Comparing $\psi(q,R;z_{\rm int})$ with $\psi(q,R;z_{\rm liq})$, and
$\phi_{\rm neut}(R;z_{\rm int})$ with $\phi_{\rm neut}(R;z_{\rm liq})$,
gives a sense for features of solvation that most strongly shape ion
adsorption. Similarities point to aspects of solvent structure and
response which are largely unchanged when an ion moves to the
interface. These contributions may be important for solvation in an
absolute sense, but their cancellation indicates a weak net influence
on adsorption thermodynamics.

For all values of $R$ we considered, $\phi_{\rm neut}$ is less
negative at $z_{\rm int}$ than at $z_{\rm liq}$.  In the simplest
conception of the liquid's boundary as a layer of nonzero dipole
density, one would expect the nonlocal component of $\phi_{\rm neut}$
to attenuate steadily in magnitude as a solute moves from the liquid
phase into the interfacial region, and then vanish as the solute
enters vapor. Whether this rough picture is consistent with the
observed shift in $\phi_{\rm neut}$ depends on the sign of the
nonlocal potential $\phi_{\rm d}$. Unfortunately this sign is
uncertain, as described in Sec.~\ref{sec:BulkReview}, due to the
intrinsic ambiguity in dividing molecules between bulk and surface
regions. Ref.~\citenum{remsing2014role} calculated a positive dipole
component of the surface potential, $\phi_{\rm d}=+260$~mV.  Within
the simple continuum picture, this value suggests a downward shift in
$\phi_{\rm neut}$ as $z$ increases from $z_{\rm liq}$ to $z_{\rm
  int}$, in contrast to our simulation results. A different
partitioning scheme, however, can give $\phi_{\rm d}<0$, suggesting an
upward shift, as we observe in simulation.

Although the direction of change in $\phi_{\rm neut}$ might be
anticipated from the sign of $\phi_{\rm d}$, the magnitude of this
shift varies considerably with solute size. For $R=0.24$\,nm,
$|\phi_{\rm neut}|$ is reduced by about 15\% when the cavity is placed
at the interface. For $R=0.415$\,nm the reduction is greater than
50\%. This variation cannot arise from nonlocal biases, which are
insensitive to the size or charge of a solute.  A distinct,
macroscopically nonlocal contribution could manifest as a nonzero
asymptotic value of $ \Delta_{\rm ads}\phi_{\rm neut} = \phi_{\rm
  neut}(R;z_{\rm int})- \phi_{\rm neut}(R;z_{\rm liq})$ at
intermediate $R$; according to our data, if such a limit exists it
occurs for solutes larger than 1\,nm.

\begin{figure}[!tb]
  \centering
  \includegraphics[width=7.68cm]{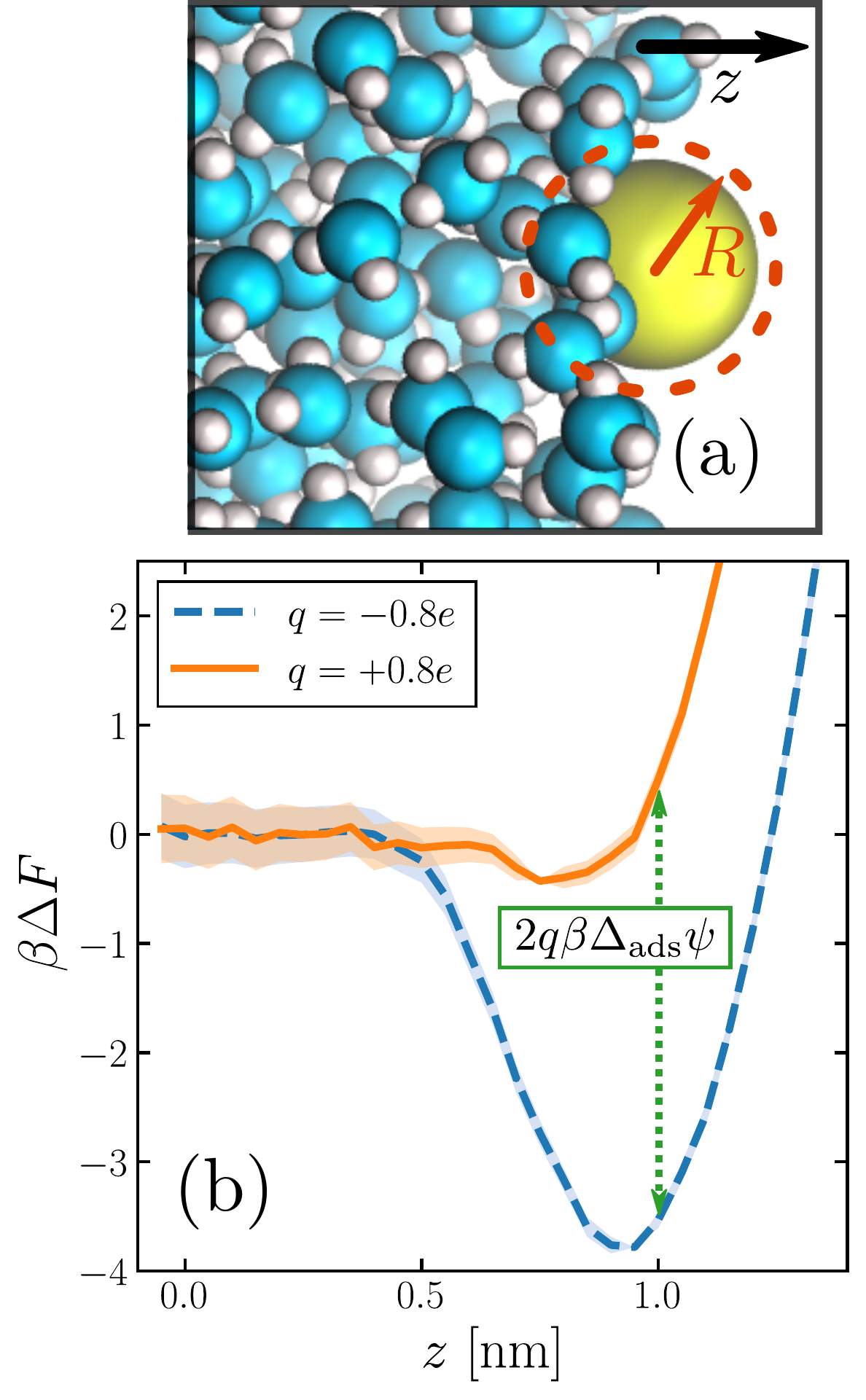}
  \caption{The propensity for an ion to adsorb to the air-water
    interface depends strongly on the sign of its charge. (a) Snapshot
    of an iodide-sized anion ($R = 0.415$\,nm) at the interface. The
    system comprises a free-standing slab of liquid water surrounded
    on either side by its vapor. (Only one of the two interfaces is
    shown.) The $z$ direction is indicated by the arrow. The size of
    the solute is depicted schematically by the dashed circle. (b)
    Potential of mean force $\Delta F$ as a function of ion position
    $z$, for a solute charge $q=+0.8e$ (solid orange) and $q=-0.8e$
    (dashed blue). The anion adsorbs much more strongly to the
    interface than the cation for this solute size.  The dotted green
    line indicates the connection between these free energy profiles
    and the adsorption asymmetry potential in
    Eq.~\ref{eqn:DeltaPsi-dens}.}
  \label{fig:pmf}
\end{figure}

\begin{figure}[!b]
  \centering
  \includegraphics[width=7.65cm]{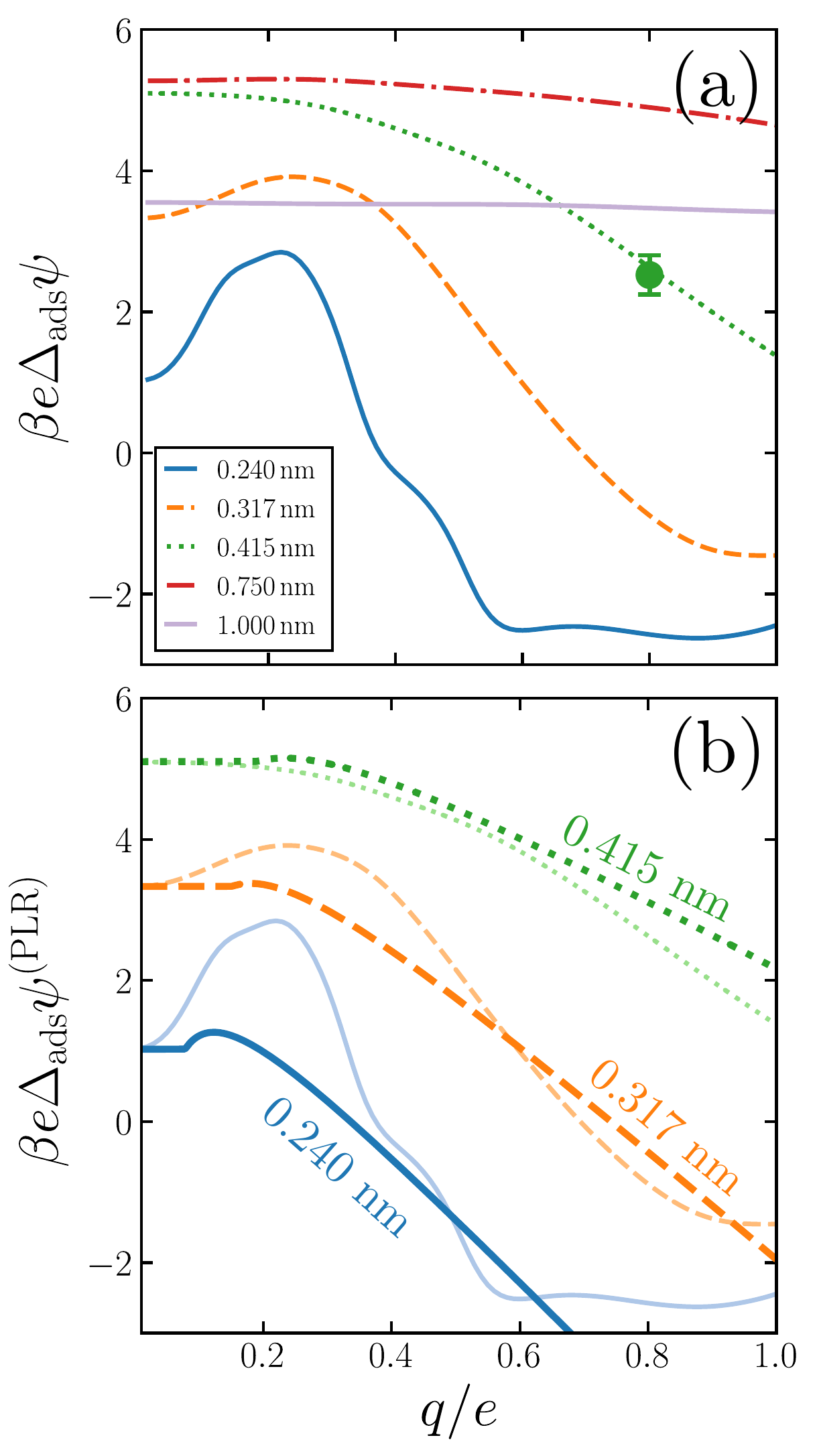}
  \caption{Linear response theory cannot faithfully describe the
    differences between adsorption profiles of sub-nanometer anions
    and cations, as demonstrated in (a) by variations in adsorption
    asymmetry potential $\Delta_{\rm ads}\psi$ with both $R$ and
    $q$. For the smallest solutes ($R \lesssim 0.4$\,nm), $\Delta_{\rm
      ads}\psi$ even changes sign as $q$ increases. In this size
    range, fully charged cations are more abundant at the interface
    than anions (with the same bulk concentration). At larger $R$,
    solutes with $q=-e$ adsorb more strongly than those with
    $q=+e$. As the solute diameter approaches $R=1$\,nm, nonlinear
    response during the charging process becomes much less
    pronounced. Values of $R$ are indicated in the legend. (b) A PLR
    model (heavy lines) predicts $\Delta_{\rm ads}\psi$ is initially
    flat, followed by a steady decrease as $q$ increases. This
    qualitatively captures the simulation data (light lines), although
    it fails to capture the leveling off at large $q$ seen for
    $R=0.240$\,nm and $0.317$\,nm.}
  \label{fig:DeltaPsis}
\end{figure}

The similarity between the asymmetry potentials $\psi(q,R)$ for
solutes in the bulk and at the interface offers some hope that
complicating factors of nonlinear response cancel out in the
adsorption process. The extent of this cancellation is quantified by
an adsorption asymmetry potential
\begin{align}
  \Delta_{\rm ads}\psi(q,R) &= \psi(q,R;z_{\rm int})
  - \psi(q,R;z_{\rm liq}), \label{eqn:DeltaPsi} \\[7pt]
  &= (2\beta q)^{-1}\ln{
    \left[{\rho_{\rm int}(-q,R;\rho_{\rm bulk})
        \over 
        \rho_{\rm int}(+q,R;\rho_{\rm bulk})}\right]}
  \label{eqn:DeltaPsi-dens},
\end{align}
where $\rho_{\rm int}$ is the average number density of a solute at
$z=z_{\rm int}$, given its concentration $\rho_{\rm bulk}$ in bulk
solution.  Eq.~\ref{eqn:DeltaPsi-dens} highlights the direct
relationship between $\Delta_{\rm ads}\psi(q,R)$ and the relative
adsorption propensities of cations and anions: For dilute solutes with
opposite charge, equal size, and equal bulk concentration,
$\exp{[2\beta q \Delta_{\rm ads}\psi(q,R)]}$ directly indicates the
enhancement of anions over cations at the interface, as shown in
Figs.~\ref{fig:pmf} and~\ref{fig:DeltaPsis}. From the preceding
discussion of the asymmetry potential itself, it is clear that
$\Delta_{\rm ads}\psi(q\to 0,R) = \Delta_{\rm ads}\phi_{\rm neut}$.
The full dependence of $\Delta_{\rm ads}\psi$ on $q$ thus incorporates
the adsorption behavior of the neutral cavity potential as well as the
corresponding solvent response to charging. Our numerical results for
$\Delta_{\rm ads}\psi(q,R)$ are the central contribution of this
paper.

The adsorption asymmetry potential $\Delta_{\rm ads}\psi(q,R)$, as
determined from simulations of the SPC/E model, are plotted as a
function of $q$ in Fig.~\ref{fig:DeltaPsis} for several values of $R$.
For the smaller solutes, the scale on which $\Delta_{\rm ads}\psi$
changes upon charging is dramatically smaller than the asymmetry
potentials themselves. Nonlinear solvent response in these cases
cancels substantially in the process of adsorption, but by no means
completely. Despite the partial cancellation, $\Delta_{\rm ads}\psi$
still varies by more than 100\,mV as $q$ increases from 0 to $e$,
comparable in magnitude to $\phi_{\rm d}$ and $\phi_{\rm neut}$.  For
$R=0.24$\,nm and $R=0.317$\,nm, this variation is sufficient to change
even the {\em sign} of $\Delta_{\rm ads}\psi$, and therefore to change
the sense of charge bias: Small monovalent cations ``adsorb'' more
favorably to the air-water interface than do anions of the same size.
In this size range, however, the adsorbed state is unstable relative
to the fully solvated ion in bulk solvent unless $q$ is very small in
magnitude.

As was previously observed for bulk solvation, we find that the
response to charging a solute at the air-water interface, while
nonlinear on the whole, is roughly piecewise linear.  Deviations from
piecewise linearity are generally stronger in the interfacial case. It
is therefore less straightforward to parameterize an interfacial
piecewise linear response model, i.e., to identify a crossover charge
$q_{\rm c}$ at which the susceptibility $d\<\phi_{\rm solv} \>_q/dq$
changes discontinuously. The SI presents plausible choices for $q_{\rm
  c}$ and these limiting susceptibilities for our three smallest
solutes, from which adsorption asymmetry potentials $\Delta_{\rm
  ads}\psi^{\rm (PLR)}$ can be readily computed. The resulting PLR
predictions are plotted in Fig.~\ref{fig:DeltaPsis}b. Two basic
features of our simulation results are accurately captured by this
phenomenological description. Specifically, (i) for small solute
charge, $\Delta_{\rm ads}\psi$ is an approximately constant or
modestly increasing function of $q$, and (ii) a more strongly
decreasing trend of $\Delta_{\rm ads}\psi$ follows for larger
$q$. Nearly quantitative agreement can be obtained for an iodide-sized
solute, $R = 0.415$\,nm.  Smaller solutes exhibit a more complicated
charge dependence that lies beyond a simple PLR description.  We note
that this test of PLR is a demanding one, given the small scale of
$\Delta_{\rm ads}\psi$ relative to $\psi(q,R;z_{\rm int})$ and
$\psi(q,R;z_{\rm liq})$ individually.  To the extent that PLR is a
successful caricature, these results suggest that the adsorption
charge asymmetry at full charging ($q=e$) derives from a combination
of features of solvent response, including an interface-induced shift
in the crossover charge $q_{\rm c}$ at which the character of linear
response changes.  The neutral cavity potential $\phi_{\rm neut}$
figures into this combination as well, but by no means does it
dominate for these solute sizes.

For the larger solutes we examined, the nonlinearity of solvent
response to charging is not pronounced, either in bulk liquid or at
the interface. The difference in nonlinearity of these environments is
necessarily also not large, with $e \Delta_{\rm ads}\psi$ changing by
less than $k_{\rm B}T$ over the range $q=0$ to $q=e$. This small
variation is comparable in scale to those of $\psi(q,R;z_{\rm int})$
and $\psi(q,R;z_{\rm liq})$ themselves. Judged on that scale, the
cancellation of nonlinear response is in fact less complete for $R =
0.75$\,nm than for smaller solutes.  As we have discussed, cavities
with $R \gtrsim 1$~nm depress the instantaneous interface
significantly, effectively placing them in the vapor phase even when
$z$ coincides with the Gibbs dividing surface.  When such a solute is
endowed with sufficient charge, wetting will occur at its surface,
eventually raising the interface to effectively move the solute into
the liquid phase. This solvent response, which originates in the
physics of phase separation, is intrinsically nonlinear. For large
$R$, a solute charge well in excess of $e$ is required to fully induce
this structural change, but at the nanoscale it may manifest as an
incipient nonlinearity for $q\approx e$.

In summary, the adsorption asymmetry potential $\Delta_{\rm ads}\psi$
depends significantly on solute size $R$ and charge $q$. Neither of
these sensitivies can arise from intrinsic orientational bias at the
neat air-water interface. Long-range electrostatic forces from
oriented molecules at the liquid's boundary, which contribute
importantly to surface potentials like $\phi_{\rm surf}$ and
$\phi_{\rm d}$, are inherently unaffected by the presence, size, or
charge of a sufficiently distant solute. These results highlight the
importance of local solvent structure and response for charge
asymmetry in interfacial ion adsorption, and they highlight the danger
of inferring solvation thermodynamics from ion-free quantities such as
$\phi_{\rm surf}$ and $\phi_{\rm d}$.

\section{Discussion and Conclusions}

In this study, we set out to understand whether or not charge
asymmetry in interfacial ion adsorption could be understood in terms
of macroscopically long-ranged, collective forces intrinsic to water.
For ion solvation in bulk, difficulties in unambiguously determining
such long-ranged contributions were already apparent from previous
results.  Building on that work, our results show that for SPC models
of water such a simple mechanistic picture is inadequate for
interfacial solvation as well. In addition to the difficulties in
partitioning molecules between `near' and `distant' interfaces,
complex nonlinear response also underlies substantial shortcomings of
trying to rationalize ion adsorption from surface potentials that
characterize biases of the undisturbed air-water interface. The
nonlinearities in $F_{\rm chg}(q,R)$ for bulk and interfacial
environments, while similar, are sufficiently different that the
process of adsorption is also substantially nonlinear. A compelling
inference of adsorption tendencies from intrinsic properties of the
undisturbed liquid and its interface with vapor requires information
that is more subtle than an average electric potential and macroscopic
dielectric susceptibility. As highlighted by the potential
distribution theorem,\cite{widom1982potential} this information can in
principle be gleaned from equilibrium statistics of the undisturbed
solvent. But in terms of fluctuations in electric potential, it
involves high-order correlations whose physical meaning is not
transparent.

In previous work we developed and tested finite size corrections for
computer simulations of interfacial ion
solvation.\cite{cox2018interfacial} Based on DCT, these corrections
proved to be quite accurate even for simulation unit cells with
nanometer dimensions. Our conclusion that DCT is a faithful
representation of aqueous polarization response down to nanometer
length scales is reinforced by the results of this paper. In
particular, when charging a solute of diameter $R=1$\,nm, solvent
response on an absolute scale is linear to a very good approximation,
both in bulk liquid and at the interface. The results of
Figs.~\ref{fig:psis} and~\ref{fig:DeltaPsis}, however, also indicate
that 1\,nm marks the validity limit of linear response. When charging
a cavity with $R=0.75$\,nm, nonlinear contributions to charge
asymmetry are quantitatively important; for smaller solutes such
nonlinear contributions become not just important but instead
dominant. In passing, we note that even for the larger solutes, a
significant charge asymmetry persists, both for bulk solvation and for
adsorption to the interface. This persistent bias weighs against the
basis of the tetra-phenyl arsonium/tetra-phenyl borate (`TATB')
extrathermodynamic assumption, an issue that has also been raised by
others.\cite{duignan2018understanding,pollard2018re,scheu2014charge,remsing2019influence}

The highly simplified description of molecular interactions in SPC
models is certainly a crude approximation to real microscopic
forces. But the specific ion effects it exhibits cannot be ascribed
simply to an errant surface potential. Indeed, discrepancies between
models in potentials such as $\phi_{\rm d}$ (whose definition requires
an arbitrary convention), $\phi_{\rm surf}$ (which pertains to a
solute that does not exclude volume), or even $\phi_{\rm neut}$ (which
for subnanometer solutes does not account for the strong asymmetry of
solvent response) are not greatly alarming. $\phi_{\rm surf}$ and
$\phi_{\rm d}$ can vary significantly among different models, but they
do not weigh on ion solvation thermodynamics in a direct way, either
in bulk liquid or at the air-water interface. (This does not
contradict their use for computing $F_{\rm chg}$ once a choice for
partitioning molecules between the interface and bulk has been made.)
By contrast, trends in $F_{\rm solv}$ and $\Delta_{\rm ads}\psi$ at
full charging reflect on essential microscopic mechanisms that
underlie specific ion adsorption. SPC models may be best viewed as
caricatures of a disordered tetrahedral network, with intrinsic charge
asymmetry due to the distinct geometric requirements of donating and
accepting hydrogen bonds. These essential features of liquid water are
often associated with nonlinear response in
solvation.\cite{maroncelli1988computer,skaf1996molecular,geissler2000importance}
By implicating nonlinearities of precisely this kind as sources of
ion-specific adsorption properties, our results support the use of SPC
models as a physically motivated test bed for exploring the
microscopic basis of surprising trends in interfacial
solvation. Conversely, our results underscore the limitations of DCT
and notions of long-ranged contributions from unperturbed interfaces,
which do not describe essential local aspects of the chemical physics
underlying ion adsorption and its charge asymmetry.  The consequences
of this shortcoming are likely to be exacerbated in confined
geometries. Work to move beyond standard DCT approaches is an active
area of research
(e.g. Refs. \citenum{loche2018breakdown,loche2020universal,schlaich2016water,zhang2020electromechanics})
and it is hoped that the results presented in this study will help to
guide future theoretical developments.

\section{Methods}

All simulations used the SPC/E water
model;\cite{BerendsenStraatsma1987sjc} solutes were represented as
Lennard-Jones (LJ) spheres with a central charge $q$. The SHAKE
algorithm was used to maintain a rigid water
geometry.\cite{ryckaert1977numerical} Periodic boundary conditions
were imposed in all three Cartesian directions, with the liquid phase
spanning two directions in a slab geometry.  Long-range Coulomb
interactions were summed using the particle-particle particle-mesh
Ewald method.\cite{HockneyEastwood1988sjc,kolafa1992cutoff} A
spatially homogeneous background charge was included to maintain
electroneutrality and thus guarantee finite electrostatic energy. For
solute sizes $R<0.75$\,nm, the system comprised 266 water molecules
with simulation cell dimensions $2\times 2\times 4.5$\,nm$^3$. For
$R\ge 0.75$\,nm the simulation cell size was $3.5\times 3.5\times
8.5$\,nm$^3$ and we used 1429 water molecules. Solvent density
profiles that indicate the interfacial location $z_{\rm int}$ for each
solute are given in the SI. A time step of 1\,fs was used for all
simulations. A temperature of 298\,K was maintained using Langevin
dynamics,\cite{DunwegWolfgang1991sjc,SchneiderStoll1978sjc} as
implemented in the LAMMPS simulation package,\cite{plimpton1995sjc}
which was used throughout.

Due to the long range of Coulomb interactions, ion solvation in polar
solvents has important contributions even from distant solvent
molecules. Thermodynamic estimates from molecular simulations are thus
subject to substantial finite size effects, which have been the focus
of many
studies.\cite{HummerGarcia1996sjc,figueirido1997finite,hummer1997ion,HunenbergerMcCammon1999sjc,HummerGarcia1998sjc}
In Ref.~\citenum{cox2018interfacial} we showed for liquid water in a
periodic slab geometry that values of $\phi_{\rm neut}$ depend on
simulation box size in a slowly decaying but predictable way. The
limit of infinitely separated periodic images can thus be obtained
with a simple finite size correction, which amounts to referencing
electric potential values to the vapor phase. We have applied this
correction to all potentials reported in this paper.  The potential of
mean force $\Delta F(q,R;z)$ for ions in periodic liquid slabs are, by
contrast, nearly independent of simulation cell size for $z\leq z_{\rm
  int}$.\cite{cox2018interfacial}

To compute $\Delta F(q,R;z)$, we followed the same procedure as
outlined in Ref.~\citenum{mccaffrey2017mechanism}, namely umbrella
sampling and histogram reweighting with MBAR.\cite{shirts2007accurate}
To calculate $\psi(q,R;z)$ for a given choice of $R$ and $z$,
simulations were performed with $q/e = -1.0, -0.9,\ldots,+0.9,
\textrm{ and } +1.0$. This spacing of $q$ values allows for ample
overlap among probability distributions $P_q\big(\phi_{\rm solv}\big)$
of the electrostatic potential at the center of the solute
(\tcr{Fig.~S10}). For $R\le0.415$\,nm statistics were obtained from
trajectories 5\,ns in duration. For $R=0.75$\,nm and $1.0$\,nm,
trajectories varied between $2.8$\,ns and $5.0$\,ns. Using the MBAR
algorithm, results from the entire range of solute charge were
combined to determine the neutral cavity distribution $P_0(\phi_{\rm
  solv})$ over a correspondingly broad range of $\phi_{\rm
  solv}$. $F_{\rm chg}$ was computed by averaging $\exp(-\beta
q\phi_{\rm solv})$ according to the distribution $P_0(\phi_{\rm
  solv})$, as prescribed by Widom's potential distribution
theorem,\cite{widom1982potential}
\begin{equation}
  \label{eqn:reweight}
e^{-\beta F_{\rm chg}}
  =
  \int\!\mrm{d}\phi_{\rm solv}\,
  P_0\big(\phi_{\rm solv}\big)
  e^{-\beta q\phi_{\rm solv}}
\end{equation}
The integral in Eq.~\ref{eqn:reweight} was performed numerically.

\acknowledgments{S.J.C (02/15 to 09/17), D.G.T, P.R.S and P.L.G were
  supported by the U.S. Department of Energy, Office of Basic Energy
  Sciences, through the Chemical Sciences Division (CSD) of Lawrence
  Berkeley National Laboratory (LBNL), under Contract
  DE-AC02-05CH11231. Since 10/17, S.J.C. has been supported by a Royal
  Commission for the Exhibition of 1851 Research Fellowship.}

\bibliography{../ver2/cox.bib}

\clearpage
\onecolumngrid
\renewcommand\thefigure{S\arabic{figure}}
\renewcommand\theequation{S\arabic{equation}}
\renewcommand\thesection{S\arabic{section}}
\setcounter{figure}{0}
\setcounter{equation}{0}
\setcounter{section}{0}

\noindent {\Large \tbf{Supporting Information}}

\vspace{0.25cm}

\noindent This document contains a detailed account of the issues
faced when trying to isolate contributions to $\phi_{\rm neut}$ from
local and distant sources. Also given are solvent density profiles
$\rho(z)$ in the presence of the neutral solute for the different
systems studied, and the position of the solute at the interface is
indicated in each instance. Solute-solvent radial distribution
functions $g(r)$ are shown for $q = -e$, $0$ and~$+e$ with the solute
in the center of the slab. Details underlying the piecewise linear
response model are also presented. A brief description of how
$P_0(\phi_{\rm solv})$ is obtained is given.

\section{Electrostatic contributions from near and far}

The challenge of identifying and interpreting a potential drop across
the liquid-vapor interface can be viewed as an issue of partitioning
molecules between distinct regions of space.

\begin{figure}[bht]
  \centerline{\includegraphics[width=15cm]{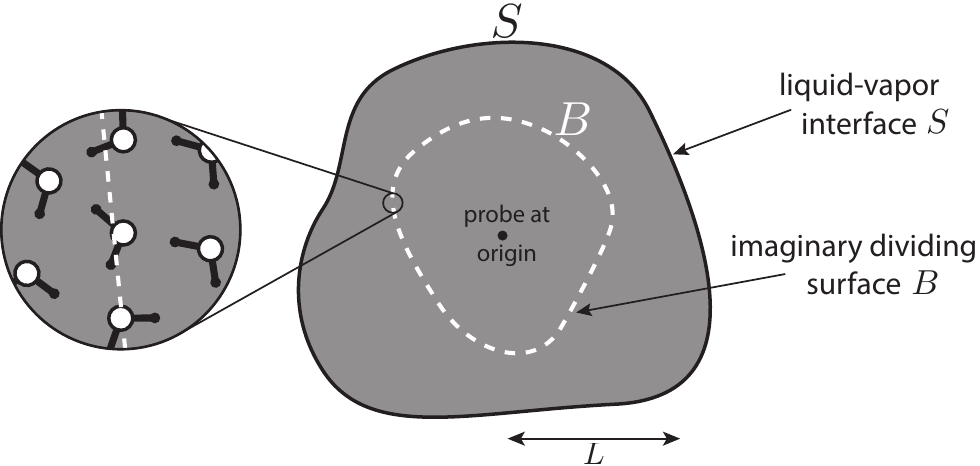}}
  \caption{Sketch of a macroscopic droplet of liquid (shaded region at
    right) surrounded by dilute vapor.  The two phases contact at a
    macroscopically smooth interface $S$.  The surface $B$ within the
    droplet is a mathematical device to isolate the electrostatic
    contribution of molecules residing near the phase boundary
    $S$. The droplet's overall scale $L$ is a macroscopic distance.  A
    magnified view of a microscopic region straddling $B$ is shown at
    left. Molecules intersected by $B$ (dashed white line) could
    reasonably be assigned to either the near (inside $B$) or far
    (outside $B$) domains.}
\label{fig:droplet}
\end{figure}

Consider a macroscopic droplet of liquid bounded by an interface $S$
with the vapor phase (as illustrated in Fig.~\ref{fig:droplet}). The
origin of our coordinate system lies deep within the bulk liquid
phase. We will aim to calculate the average electric potential $\<\phi
\>$ at the origin, distinguishing contributions of molecules that are
far from the probe (including those at the phase boundary) from those
that lie nearer the origin. Specifically, we will divide the two
populations at an imaginary surface $B$ that is also deep within the
bulk liquid.  We will take $B$ to be distant enough from the origin
that liquid structure on this surface is bulk in character, even if
the microscopic vicinity of the origin is complicated by a solute's
excluded volume.

\subsection{Partitioning schemes}

The vast majority of molecules in the droplet are unambiguously
located either outside $B$ (``far'') or inside $B$ (``near'').  A tiny
fraction straddle the surface $B$. In the case of water this could
involve a molecule's oxygen atom lying on one side of $B$, while its
hydrogen atoms lie on the other. One division scheme (an M-scheme)
would judge the molecule's location based on the O atom; another
M-scheme might base the classification on the molecule's center of
charge. A still different scheme (a P-scheme) could divide the
molecule in two, with some pieces ``near'' and other pieces
``far''. (The M-scheme and the P-scheme are well known in the
literature. See e.g. Ref.~\onlinecite{kastenholz2006computation}.) The
total potential $\phi$ at the probe site is not sensitive to which of
these schemes is chosen. But its contributions $\phi_\trm{near}$ and
$\phi_\trm{far}$ from atoms/molecules in the near and far regions {\em
  are} sensitive, in an offsetting way.

Let's first treat the M scheme, with the molecule's near/far
classification based on the position $\mbf{r}^{(0)}$ of some site
within the molecule (say, its O atom). The average far-field potential
in this case is
\beq
\< \phi_\trm{far}^\trm{M} \> = N \int_\trm{outside $B$} d\mbf{r} \int d\mbf{\Omega}\,
p(\mbf{r},\mbf{\Omega})
\sum_\alpha {q_\alpha \over |\mbf{r} + \Delta\mbf{r}_\alpha(\mbf{\Omega})|},
\eeq
where $N$ is the total number of molecules in the droplet and $\alpha$
indexes charged sites within each molecule. Here,
$p(\mbf{r},\mbf{\Omega}) = \langle
\delta(\mbf{r}-\mbf{r}^{(0)})\delta(\mbf{\Omega}-\mbf{\Omega}^{(0)})\rangle$
is the joint probability distribution of a molecule's position (i.e.,
$\mbf{r}^{(0)}$) and intramolecular configuration $\mbf{\Omega}^{(0)}$
(specified relative to the reference position $\mbf{r}^{(0)}$, as
indicated by the superscript).
\footnote{For a rigid molecule $\mbf{\Omega}^{(0)}$ is defined simply
  by the molecule's orientation, e.g., a set of Euler angles.} By
$\Delta\mbf{r}_\alpha = \mbf{r}_\alpha - \mbf{r}^{(0)}$ we denote the
displacement of charge $q_\alpha$ from the reference point
$\mbf{r}^{(0)}$. This intramolecular displacement is entirely
determined by $\mbf{\Omega}^{(0)}$.

For the P-scheme, each charge $\alpha$ contributes to
$\phi_\trm{far}^\trm{P}$ if $\mbf{r}_\alpha$ lies outside $B$. The
corresponding far-field potential is
\begin{eqnarray}
  \< \phi_\trm{far}^\trm{P} \>
&=& N \sum_\alpha \int_\trm{outside $B$} d\mbf{r}
\int d\mbf{\Omega}\,
p_\alpha(\mbf{r},\mbf{\Omega}) 
{q_\alpha \over |\mbf{r}|}
\\
  &=& N \sum_\alpha \int_\trm{outside $B$} d\mbf{r}
\int d\mbf{\Omega}\,
p(\mbf{r}-\Delta\mbf{r}_\alpha,\mbf{\Omega}) 
{q_\alpha \over |\mbf{r}|}
\label{equ:pscheme}
\end{eqnarray}
where $p_\alpha(\mbf{r},\mbf{\Omega})$ is the joint probability
distribution for site position $\mbf{r}_\alpha$ and intramolecular
configuration of a solvent molecule.  In Eq.~\ref{equ:pscheme} we
have made use of the connection
\begin{eqnarray}
p_\alpha(\mbf{r},\mbf{\Omega}) &=&
\langle \delta(\mbf{r}-\mbf{r}_\alpha)
\delta(\mbf{\Omega}-\mbf{\Omega}^{(0)})\rangle
= 
\langle \delta(\mbf{r}-\Delta\mbf{r}_\alpha-\mbf{r}^{(0)})
\delta(\mbf{\Omega}-\mbf{\Omega}^{(0)})\rangle
\\
&=& p(\mbf{r}-\Delta\mbf{r}_\alpha(\mbf{\Omega}),\mbf{\Omega})
\end{eqnarray}
\\
between the distributions $p$ and $p_\alpha$.

\subsection{Multipole expansion}
Since the entire ``far'' region is macroscopically distant from the
origin, small-$\Delta\mbf{r}_\alpha$ expansions of $|\mbf{r} +
\Delta\mbf{r}_\alpha|^{-1}$ and
$p(\mbf{r}-\Delta\mbf{r}_\alpha,\mbf{\Omega})$ are well justified.
These yield
\beq
\sum_\alpha {q_\alpha \over |\mbf{r} + \Delta\mbf{r}_\alpha|} =
\left(\sum_\alpha q_\alpha\Delta\mbf{r}_\alpha \right)\cdot\nabla{1\over r}
+ {1\over 2}\left(\sum_\alpha q_\alpha \Delta\mbf{r}_\alpha\Delta\mbf{r}_\alpha
\right)
: \nabla \nabla{1\over r}+\ldots
\eeq
and
\beq
\sum_\alpha q_\alpha p(\mbf{r}-\Delta\mbf{r}_\alpha,\mbf{\Omega}) =
-\nabla\cdot \sum_\alpha q_\alpha\Delta\mbf{r}_\alpha p(\mbf{r},\mbf{\Omega})
+{1\over 2}\nabla \nabla: \sum_\alpha q_\alpha \Delta\mbf{r}_\alpha\Delta\mbf{r}_\alpha
p(\mbf{r},\mbf{\Omega})+\ldots
\eeq
where we have omitted leading terms proportional to $\sum_\alpha
q_\alpha$, which vanish by molecular charge neutrality.  When carried
through subsequent calculations, terms beyond quadrupole order in
these expansions would vanish due either to symmetry or to the
macroscopic scale of the droplet.

Defining dipole and quadrupole densities as
\beq
\mbf{m}(\mbf{r}) = N \int d\mbf{\Omega}\,
p(\mbf{r},\mbf{\Omega})
\sum_\alpha q_\alpha\Delta\mbf{r}_\alpha 
\eeq
and
\beq
\mbf{Q}(\mbf{r}) = {N\over 2} \int d\mbf{\Omega}\,
p(\mbf{r},\mbf{\Omega})
\sum_\alpha q_\alpha \Delta\mbf{r}_\alpha\Delta\mbf{r}_\alpha
\eeq
we can write
\beq
\< \phi_\trm{far}^\trm{M} \> = \int_\trm{outside $B$} d\mbf{r}
\left(
\mbf{m}(\mbf{r})\cdot\nabla{1\over r}
+ \mbf{Q}(\mbf{r}):\nabla\nabla{1\over r}+\ldots
\right)
\label{equ:phim}
\eeq
and
\beq
\< \phi_\trm{far}^\trm{P} \> = \int_\trm{outside $B$} d\mbf{r}
{1\over r}
\left(
-\nabla\cdot\mbf{m}(\mbf{r})
+ \nabla\nabla:\mbf{Q}(\mbf{r})+\ldots
\right)
\label{equ:phip}
\eeq
Integrating by parts, and noting that $\mbf{m}(\mbf{r})$
and $\nabla:\mbf{Q}(\mbf{r})$ vanish both on $B$ and at infinity,
\begin{eqnarray}
\< \phi_\trm{far}^\trm{P} \> &=& \int_\trm{outside $B$} d\mbf{r}
\left(
\mbf{m}(\mbf{r})\cdot\nabla{1\over r}
-\left(\nabla{1\over r}\right)\cdot\big(\nabla\cdot \mbf{Q}(\mbf{r})\big)
\right)
\\
&=& 
\int_\trm{outside $B$} d\mbf{r}
\left(
\mbf{m}(\mbf{r})\cdot\nabla{1\over r}
-\nabla\cdot\left(\nabla{1\over r}\cdot\mbf{Q}(\mbf{r}) \right)
+ 
\mbf{Q}(\mbf{r}):\nabla\nabla{1\over r}
\right)
\end{eqnarray}
Using the divergence theorem,
\beq
\< \phi_\trm{far}^\trm{P} \> = 
\< \phi_\trm{far}^\trm{M} \> -
\int_B d\mbf{R} \,\, \hat{\mbf n}(\mbf{R}) \cdot \left(
\nabla {1\over R}\cdot \mbf{Q}(\mbf{R})
\right)
\eeq
where $\mbf{R}$ is a point on $B$ and
$\hat{\mbf n}(\mbf{R})$ is the corresponding local {\em inward}-pointing normal vector.
Since $B$ lies within the
bulk liquid,
where the average quadrupole density $\mbf{Q}_\trm{liq}$ is isotropic,
$\mbf{Q}(\mbf{r})=
\mbf{I} \, (\trm{Tr} \mbf{Q}_\trm{liq}/3)$ everywhere on this surface.
As a result,
\begin{eqnarray}
  \< \phi_\trm{far}^\trm{P} \> &=&
\< \phi_\trm{far}^\trm{M} \> + {\trm{Tr} \mbf{Q}_\trm{liq}\over 3}
\int_\trm{inside $B$} d\mbf{r}
\nabla^2 {1\over r}
\\
&=&
\< \phi_\trm{far}^\trm{M} \> - 
   {4\pi\over 3}\trm{Tr} \mbf{Q}_\trm{liq}
\end{eqnarray}
These two measures of the far-field potential are thus different.
Moreover, the quadrupole trace that determines this difference depends
on the choice of $\mbf{r}^{(0)}$. This ambiguity is a well-known
feature of the so-called Bethe potential $-{(4\pi/ 3)}\trm{Tr}
\mbf{Q}_\trm{liq}$
\cite{HunenbergerAndReifBook,remsing2014role,duignan2017electrostatic,kathmann2011understanding,remsing2019influence,doyle2019importance}.

\subsection{Dipole surface potential}
To simplify the result for $\< \phi_\trm{far}^\trm{M} \>$, note that
$\mbf{Q}(\mbf{r})$ is isotropic everywhere
outside $B$,
except in the microscopic
vicinity of $S$. In the bulk regions of the far domain, we then have
$\mbf{Q}(\mbf{r}):\nabla\nabla r^{-1} \propto \delta(\mbf{r})=0$.  The
final term in Eq.~\ref{equ:phim} therefore has nonzero contributions
only from a thin shell whose volume is proportional to $L^2$, where
$L$ is the macroscopic scale of the droplet. Since $\nabla\nabla
r^{-1} \sim L^{-3}$ in this shell, the quadrupolar contribution to $\<
\phi_\trm{far}^\trm{M} \>$ has a negligible magnitude, $L^{-1}$. As a
result,
\beq
\< \phi_\trm{far}^\trm{M} \> = \int_\trm{outside $B$} d\mbf{r}
\,
\mbf{m}(\mbf{r})\cdot\nabla{1\over r}
\label{equ:dip}
\eeq
This integral similarly has nonzero contributions only from a
microscopically thin shell of broken symmetry, centered on the phase
boundary $S$. Since the macroscopic surface is very smooth on this
scale, and because the average dipole density points normal to the
locally planar interface, the far-field potential may be written
\beq
\< \phi_\trm{far}^\trm{M} \> =
\int_S d\mbf{R} \int dz\, m_\perp(z)\,
\hat{\mbf n}(\mbf{R})\cdot \nabla {1\over r},
\eeq
where $\mbf{R}$ is the point on $S$ nearest to $\mbf{r}$, the
coordinate $z = (\mbf{r} - \mbf{R})\cdot \hat{\mbf n}(\mbf{R})$ is the
perpendicular displacement from the liquid-vapor interface, $\hat{\mbf
  n}(\mbf{R})$ is the outward-pointing normal of $S$, and $m_\perp(z)
\hat{\mbf n}(\mbf{R})$ is the average dipole field at
$\mbf{r}$. Neglecting contributions of $\mathcal{O}(z/L)$, we may
replace $r^{-1}$ by $R^{-1}$, and easily evaluate the surface
integral, yielding
\beq
\< \phi_\trm{far}^\trm{M} \> =
-4\pi 
\int_{z_\trm{liq}}^{z_\trm{vap}} dz \,
m_\perp(z),
\label{equ:dip-fin}
\eeq
where the integral is performed in the direction from liquid
($z_\trm{liq}<0$) to vapor ($z_\trm{vap}>0$).  For the case of a
perfectly planar interface, this result is a familiar component of the
surface potential, identified by Remsing et al. as the surface dipole
contribution \cite{remsing2014role,remsing2019influence}. As they
note, its value depends on the reference position $\mbf{r}^{(0)}$
defining the molecular reference frame. In our calculation this
dependence arises from the way molecules are classified relative to
the dividing surface $B$.

\subsection{Near-field potential}
In evaluating $\< \phi_\trm{far} \>$, we have made no assumptions
about the liquid's structure near the probe. If the origin lies inside
a solute's excluded volume, then the near-field potential is
complicated by the microscopically heterogeneous arrangement of solvent
molecules in its vicinity. If, however, the probe is simply a point within
the isotropic bulk liquid, then $\< \phi_\trm{near} \>$ can be easily
determined.

For a probe that resides in uniform bulk liquid, $\mbf{m}(\mbf{r})=0$
and $\mbf{Q}(\mbf{r})=\mbf{Q}_\trm{liq}$ everywhere inside $B$.  In
the P-scheme we can conclude immediately from the analogue of
Eq.~\ref{equ:phip} that $\< \phi_\trm{near} \>=0$. In the M-scheme we
have
\beq
\< \phi_\trm{near}^\trm{M} \> = \int_\trm{inside $B$} d\mbf{r}
\,{\trm{Tr} \mbf{Q}_\trm{liq}\over 3}\,
\nabla^2{1\over r} =
-{4\pi\over 3} \trm{Tr} \mbf{Q}_\trm{liq}
\eeq
In either case the total potential sums to
\begin{eqnarray}
\< \phi\> &=& \< \phi_\trm{near}^\trm{M} \> + \<
\phi_\trm{far}^\trm{M} \>
\\
&=&
\< \phi_\trm{near}^\trm{P} \> + \<\phi_\trm{far}^\trm{P} \> 
\\
&=& -4\pi 
\int_{z_\trm{liq}}^{z_\trm{vap}} dz \, m_\perp(z)
-{4\pi\over 3} \trm{Tr} \mbf{Q}_\trm{liq}
\end{eqnarray}

These calculations of local and nonlocal contributions to the mean
electrostatic potential resemble previous developments of surface
potential in many ways
\cite{wilson1989comment,beck2013influence,remsing2014role,duignan2017electrostatic,remsing2019influence,arslanargin2012free,doyle2019importance,horvath2013vapor}. Ours
are somewhat more general than standard calculations, in that we do
not require a specific shape of the liquid domain. (The standard
development presumes an idealized geometry of the liquid phase
e.g. planar interface or spherical droplet, and integrates the
resulting 1-dimensional Poisson equation.)  More interestingly, it
places the ambiguities surrounding surface potential in an easily
conceived context: The electrostatic bias of an interface is not well
defined because there is no unique way to assign molecules to that
interface. Any attempt to do so carries an arbitrariness that (in the
case of water) is comparable in magnitude to the apparent surface
potential itself.

\clearpage

\section{Solvent density profiles and solute-solvent radial distribution functions}

\begin{figure}[hb!]
  \includegraphics[width=7.5cm]{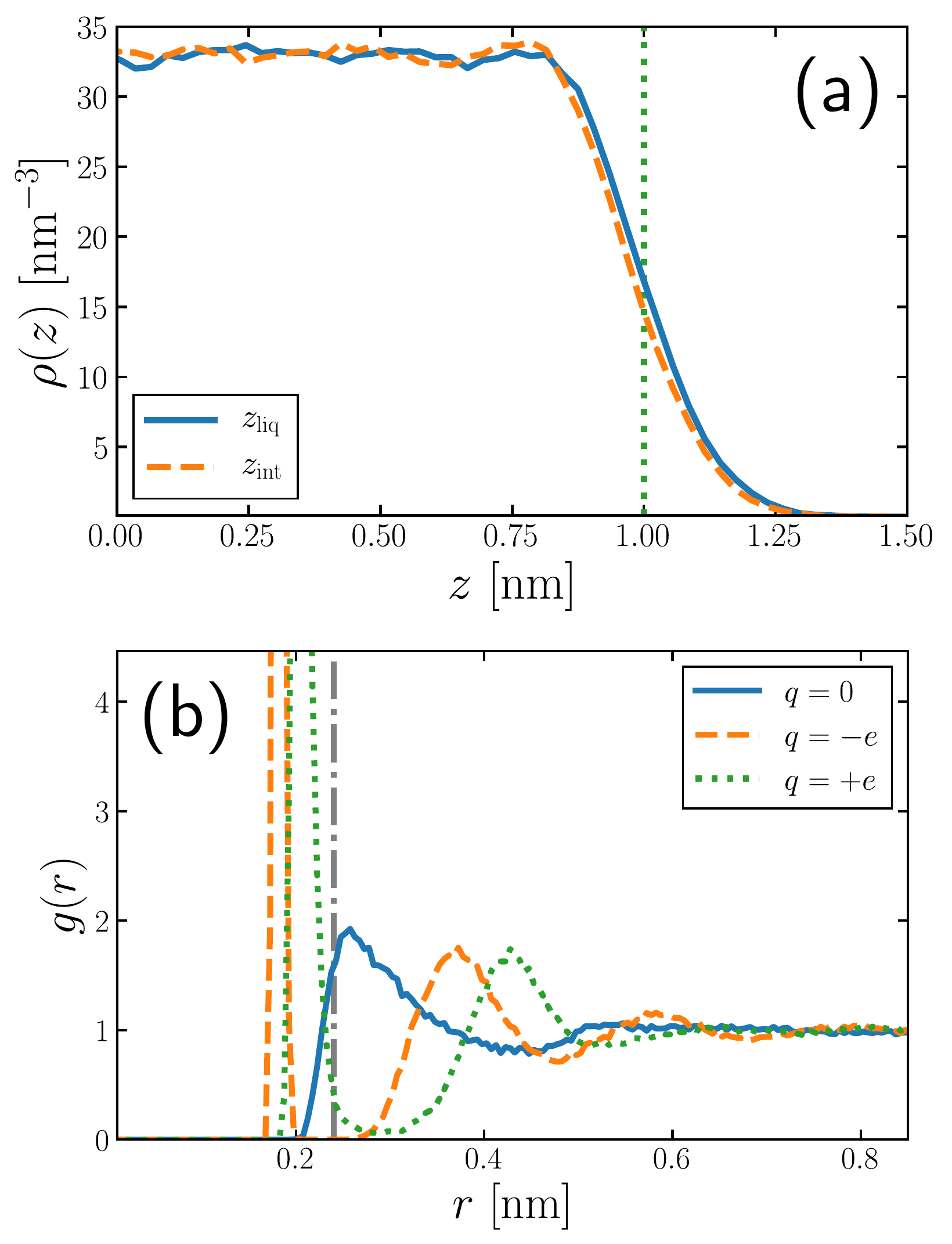}  
  \caption{(a) Average solvent density $\rho(z)$, plotted as a
    function of the coordinate $z$ perpendicular to the liquid-vapor
    interface, with the solute ($R=0.240$\,nm) located in the bulk
    ($z_{\rm liq}=0$\,nm, solid blue line), and at the interface
    ($z_{\rm int}=1$\,nm, dashed orange line). The dotted green line
    is drawn at $z=z_{\rm int}$. Only half ($z>0$\,nm) of the solvent
    profile is shown. (b) Radial distribution function $g(r)$, plotted
    as a function of the distance $r$ between the solute's center and
    the oxygen atom of a water molecule, with the solute at $z=z_{\rm
      liq}$ with $q=-e$, $0$, and~$+e$. The vertical dot-dashed gray
    line is drawn at $r=R$.}
  \label{fig:dens2o4}
\end{figure}

\begin{figure}[tb]
  \includegraphics[width=7.5cm]{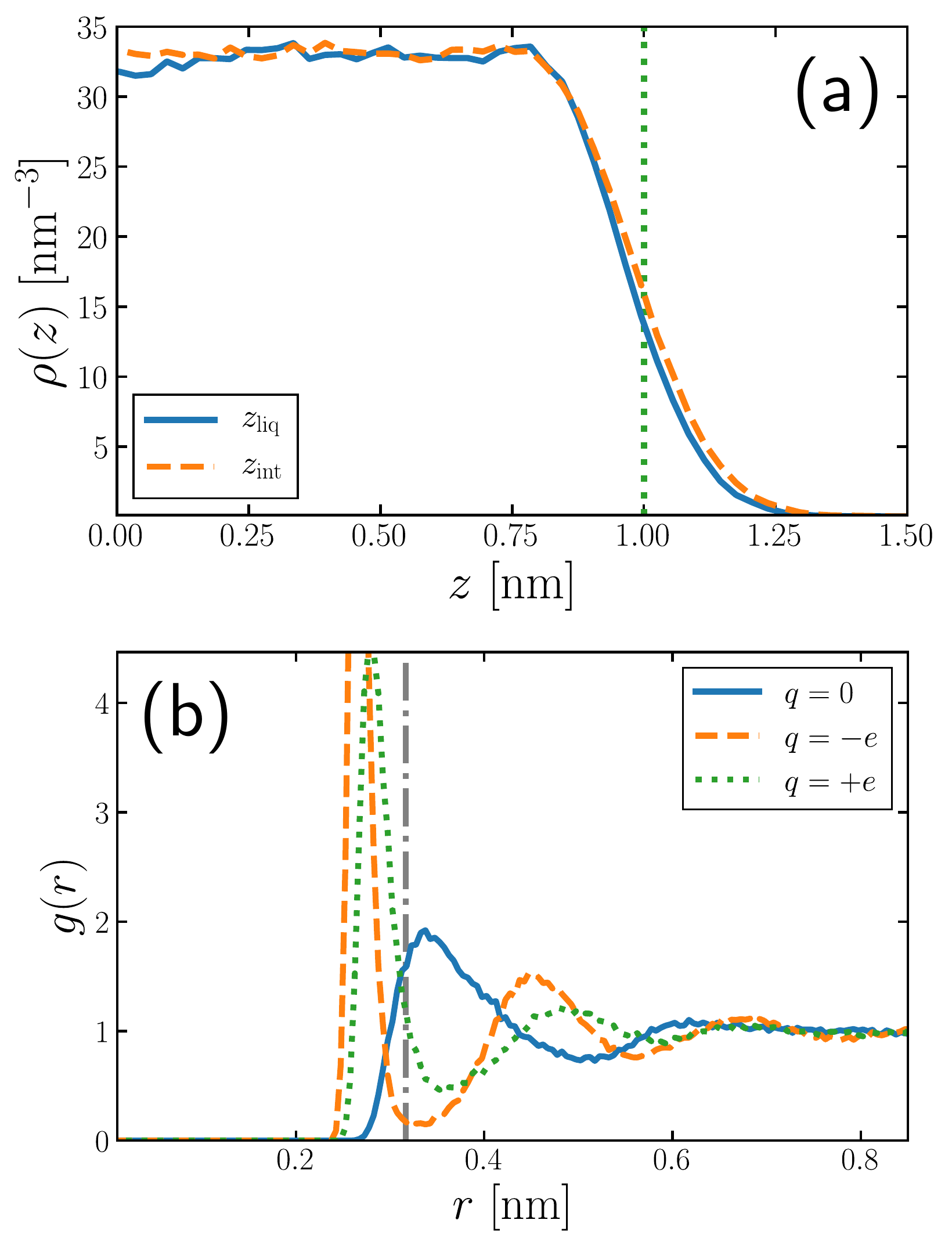}  
  \caption{(a) $\rho(z)$ with the solute ($R=0.317$\,nm) located in
    the bulk ($z_{\rm liq}=0$\,nm, solid blue line), and at the
    interface ($z_{\rm int}=1$\,nm, dashed orange line). The dotted
    green line is drawn at $z=z_{\rm int}$. Only half ($z>0$\,nm) of
    the profile is shown. (b) $g(r)$ with the solute at $z=z_{\rm
      liq}$ with $q=-e$, $0$, and~$+e$. The vertical dot-dashed gray
    line is drawn at $r=R$.}
  \label{fig:dens3o17}
\end{figure}

\begin{figure}[tb]
  \includegraphics[width=7.5cm]{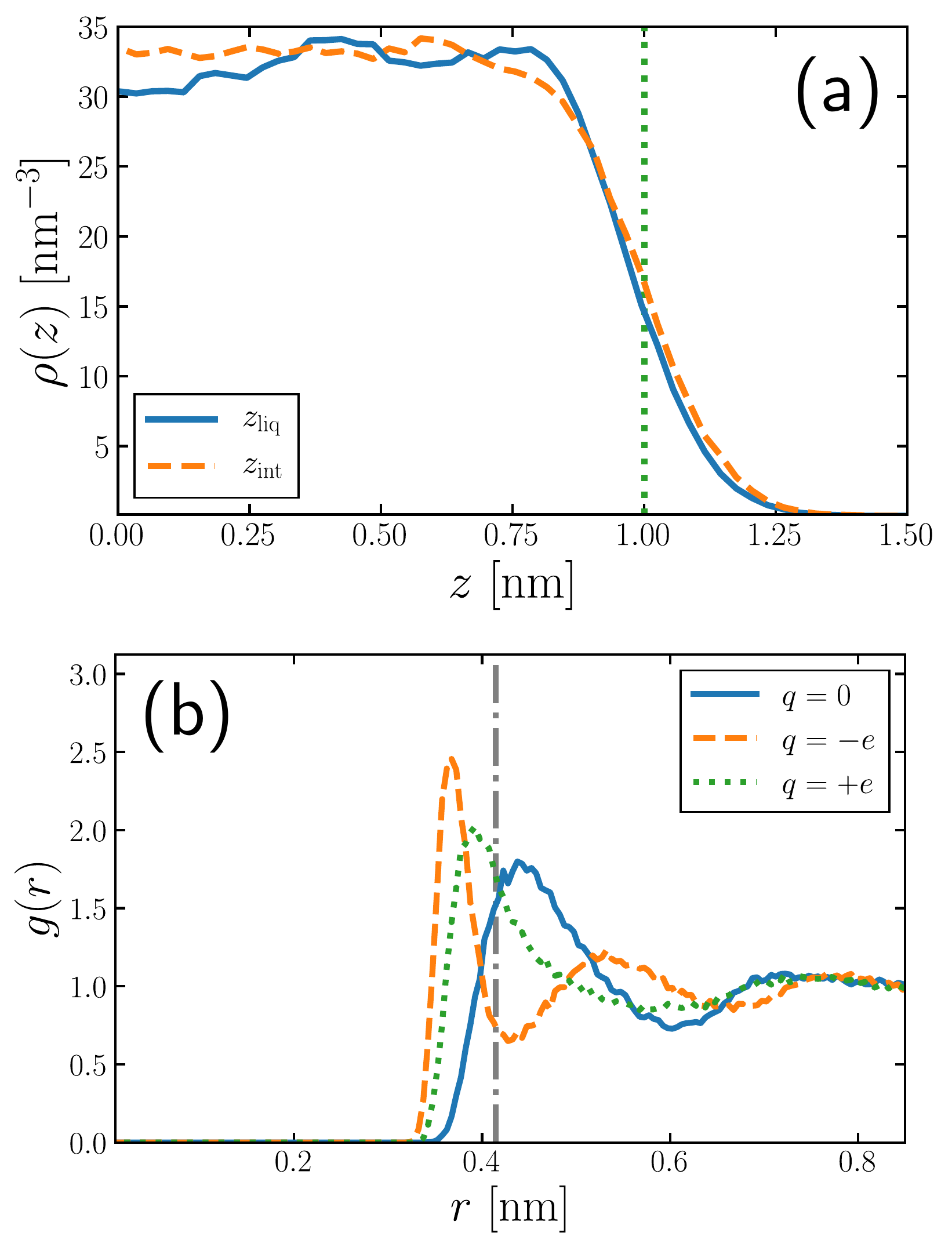}  
  \caption{(a) $\rho(z)$ with the solute ($R=0.415$\,nm) located in
    the bulk ($z_{\rm liq}=0$\,nm, solid blue line), and at the
    interface ($z_{\rm int}=1$\,nm, dashed orange line). The dotted
    green line is drawn at $z=z_{\rm int}$. Only half ($z>0$\,nm) of
    the profile is shown. (b) $g(r)$ with the solute at $z=z_{\rm
      liq}$ with $q=-e$, $0$, and~$+e$. The vertical dot-dashed gray
    line is drawn at $r=R$.}
  \label{fig:dens4o15}
\end{figure}

\begin{figure}[tb]
  \includegraphics[width=7.5cm]{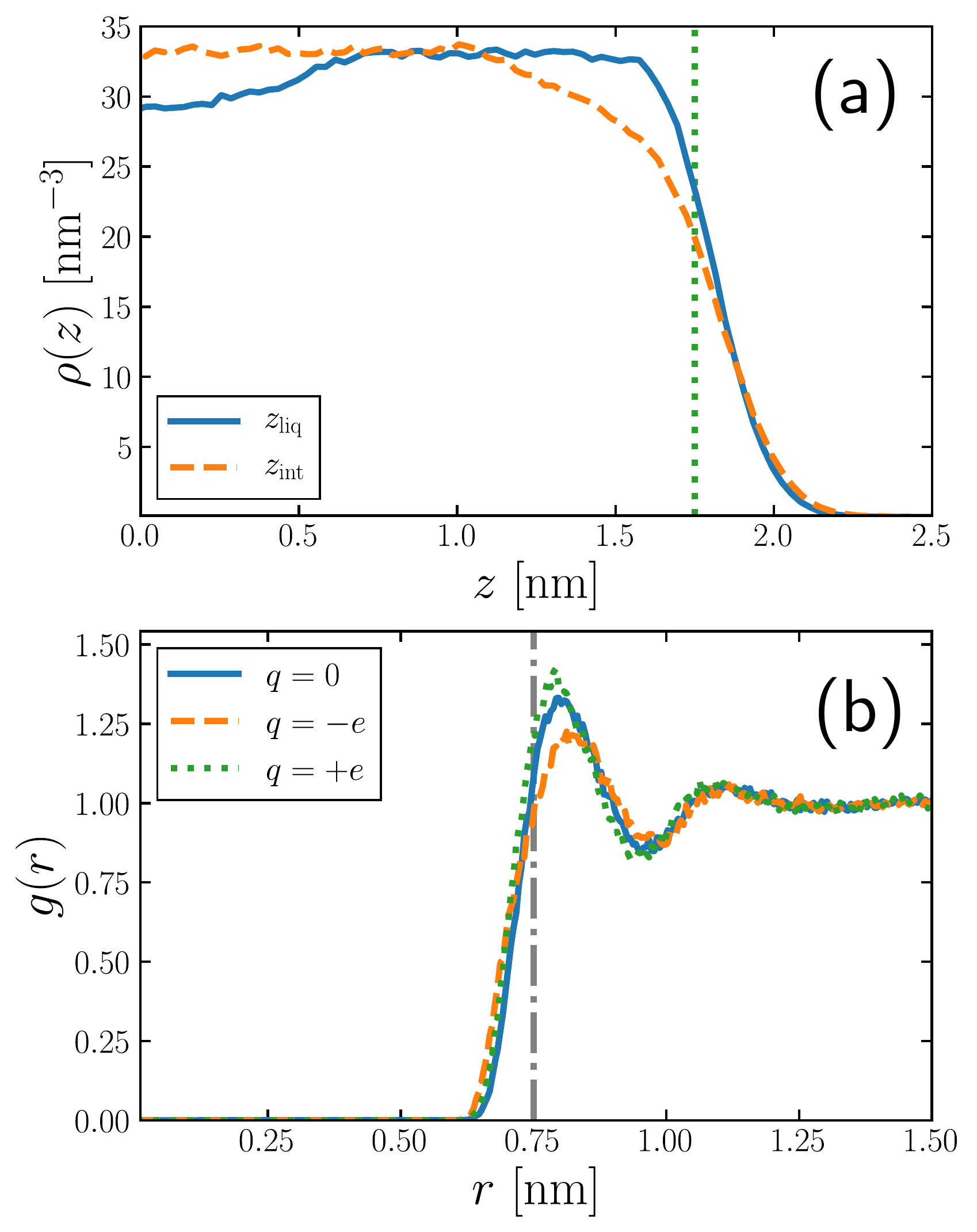}  
  \caption{(a) $\rho(z)$ with the solute ($R=0.75$\,nm) located in the
    bulk ($z_{\rm liq}=0$\,nm, solid blue line), and at the interface
    ($z_{\rm int}=1.75$\,nm, dashed orange line). The dotted green
    line is drawn at $z=z_{\rm int}$. Only half ($z>0$\,nm) of the
    profile is shown. (b) $g(r)$ with the solute at $z=z_{\rm liq}$
    with $q=-e$, $0$, and~$+e$. The vertical dot-dashed gray line is
    drawn at $r=R$.}
  \label{fig:dens7o5}
\end{figure}

\begin{figure}[tb]
  \includegraphics[width=7.5cm]{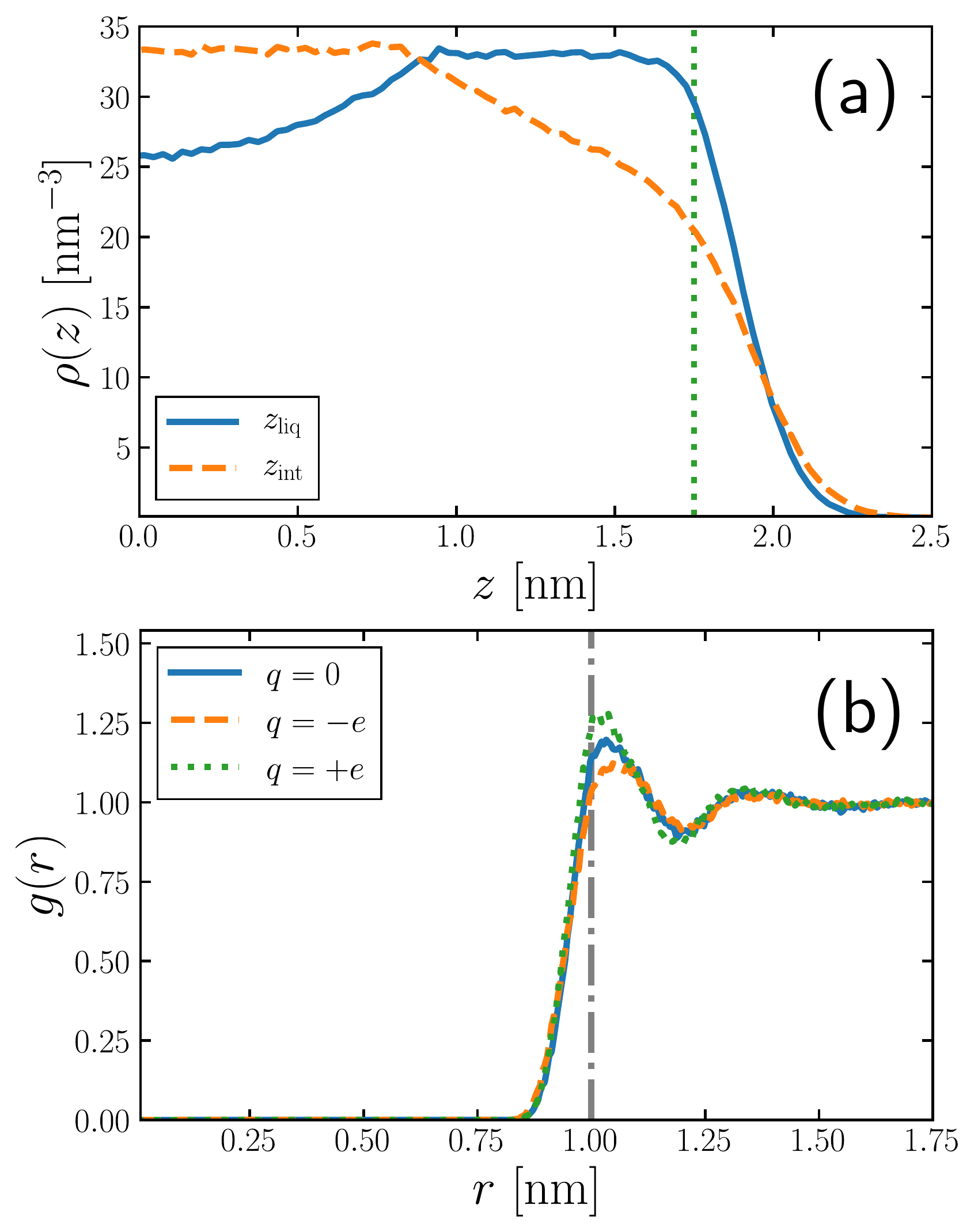}
  \caption{(a) $\rho(z)$ with the solute ($R=1.0$\,nm) located in the
    bulk ($z_{\rm liq}=0$\,nm, solid blue line), and at the interface
    ($z_{\rm int}=1.75$\,nm, dashed orange line). The dotted green
    line is drawn at $z=z_{\rm int}$. Only half ($z>0$\,nm) of the
    profile is shown. (b) $g(r)$ with the solute at $z=z_{\rm liq}$
    with $q=-e$, $0$, and~$+e$. The vertical dot-dashed gray line is
    drawn at $r=R$.}
  \label{fig:dens10o0}
\end{figure}

\clearpage

\section{Evaluating piecewise linear response}

\subsection{Outline}

Here we present details of the piecewise linear response (PLR) model
discussed in the main article. The PLR model is based on the
observation that solvent response to charging a solute is linear for
both anions and cations, but differs between the two cases
\cite{HummerGarcia1996sjc,lynden1997hydrophobic,bardhan2012affine}. In
such a model, the average electrostatic potential due to the solvent
at the center of a charged cavity can be written as
\begin{equation}
  \langle\phi_{\rm solv}\rangle_q =
  \begin{cases}
    \phi_{\rm neut} - \beta q\langle(\delta\phi_{\rm solv})^2\rangle_+\qquad&\text{($q\ge q_{\rm c}$)} \\
    \phi_{\rm neut} - \beta q\langle(\delta\phi_{\rm solv})^2\rangle_- - \beta q_{\rm c}\left[\langle(\delta\phi_{\rm solv})^2\rangle_+ - \langle(\delta\phi_{\rm solv})^2\rangle_-\right] \qquad&\text{($q<q_{\rm c}$)},
  \end{cases}
\end{equation}
where $q_{\rm c}$ is the value of the `crossover charge' between the
two linear regimes, $\langle(\delta\phi_{\rm solv})^2\rangle_+$ is the
variance of $\phi_{\rm solv}$ for $q\ge q_{\rm c}$, and
$\langle(\delta\phi_{\rm solv})^2\rangle_-$ is the variance of
$\phi_{\rm solv}$ for $q<q_{\rm c}$. (As written, it is implicitly
assumed that $q_{\rm c}\le 0$, as suggested by simulations.) Let us
define $J=\left[\langle(\delta\phi_{\rm solv})^2\rangle_+ -
  \langle(\delta\phi_{\rm solv})^2\rangle_-\right]$. $F_{\rm chg}$ is
then,
\begin{equation}
  F_{\rm chg}(q) =
  \begin{cases}
    q\phi_{\rm neut} - \frac{\beta q^2}{2}\langle(\delta\phi_{\rm solv})^2\rangle_+\qquad&\text{($q\ge q_{\rm c}$)} \\
    q\phi_{\rm neut} - \frac{\beta q^2}{2}\langle(\delta\phi_{\rm solv})^2\rangle_- -\beta J\left(qq_c - \frac{q_c^2}{2}\right)
    \qquad&\text{($q<q_{\rm c}$)},
  \end{cases}
\end{equation}
and $\psi$ is,
\begin{equation}
  \label{eqn:PLR_psi}
  \psi(q) =
  \begin{cases}
    \phi_{\rm neut}\qquad&\text{($q\le |q_{\rm c}|$)} \\
    \phi_{\rm neut}
    -\frac{\beta J}{4q}(q-|q_c|)^2 \qquad&\text{($q>|q_{\rm c}|$)}.
  \end{cases}
\end{equation}
In general, $\phi_{\rm neut}$, $q_{\rm c}$ and $J$ will depend upon
solute size, and whether or not the solute is located in bulk or at
the interface.

\subsection{Results}

\begin{figure}[tb]
  \includegraphics[width=7.5cm]{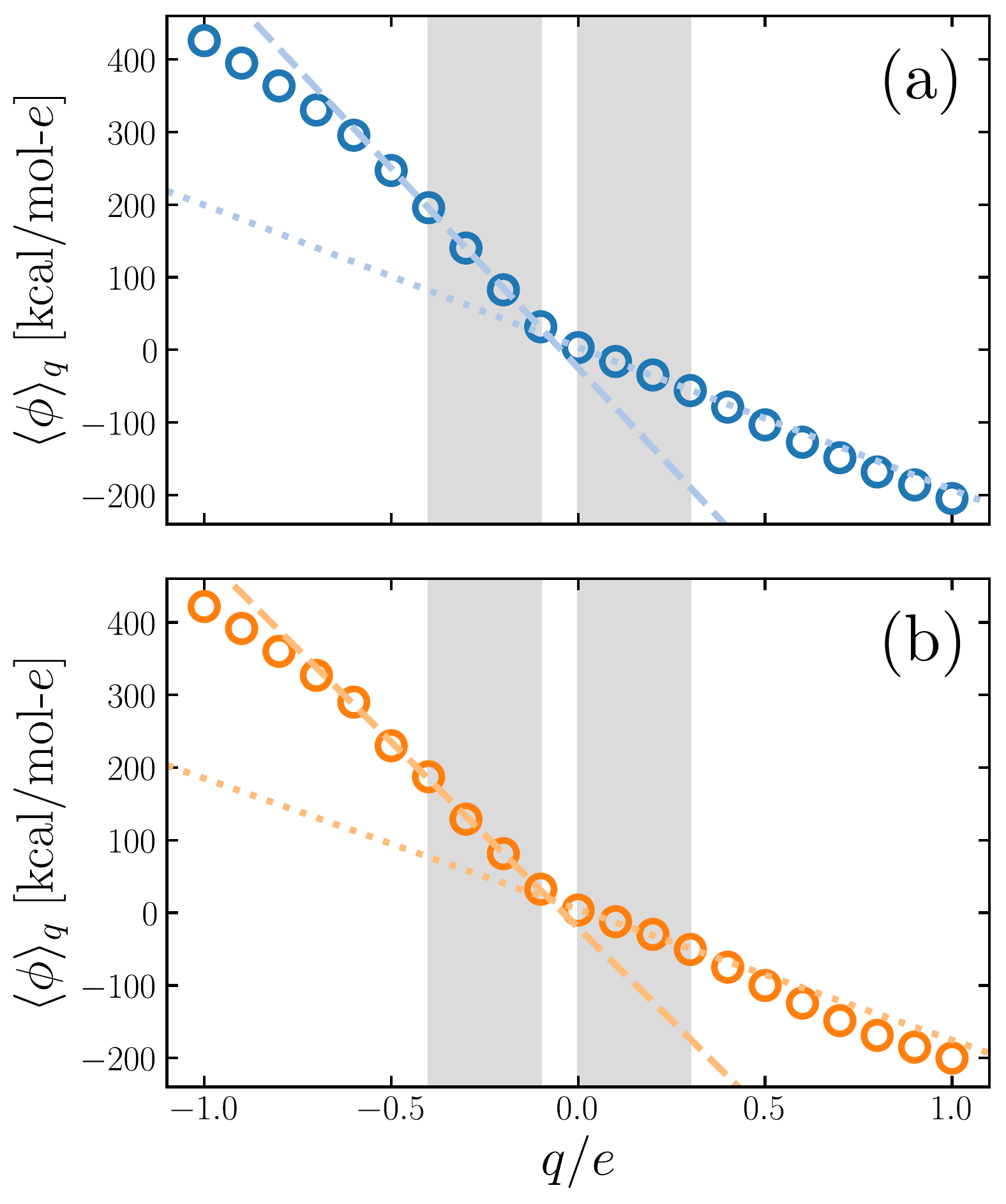}  
  \caption{$\langle\phi_{\rm solv}\rangle_q$ vs $q$ for $R=0.240$\,nm
    with the solute located (a) in bulk and (b) at the interface. The
    dashed and dotted lines show linear fits to the left and right
    shaded regions, respectively.}
  \label{fig:PLR2o4}
\end{figure}

\begin{figure}[tb]
  \includegraphics[width=7.5cm]{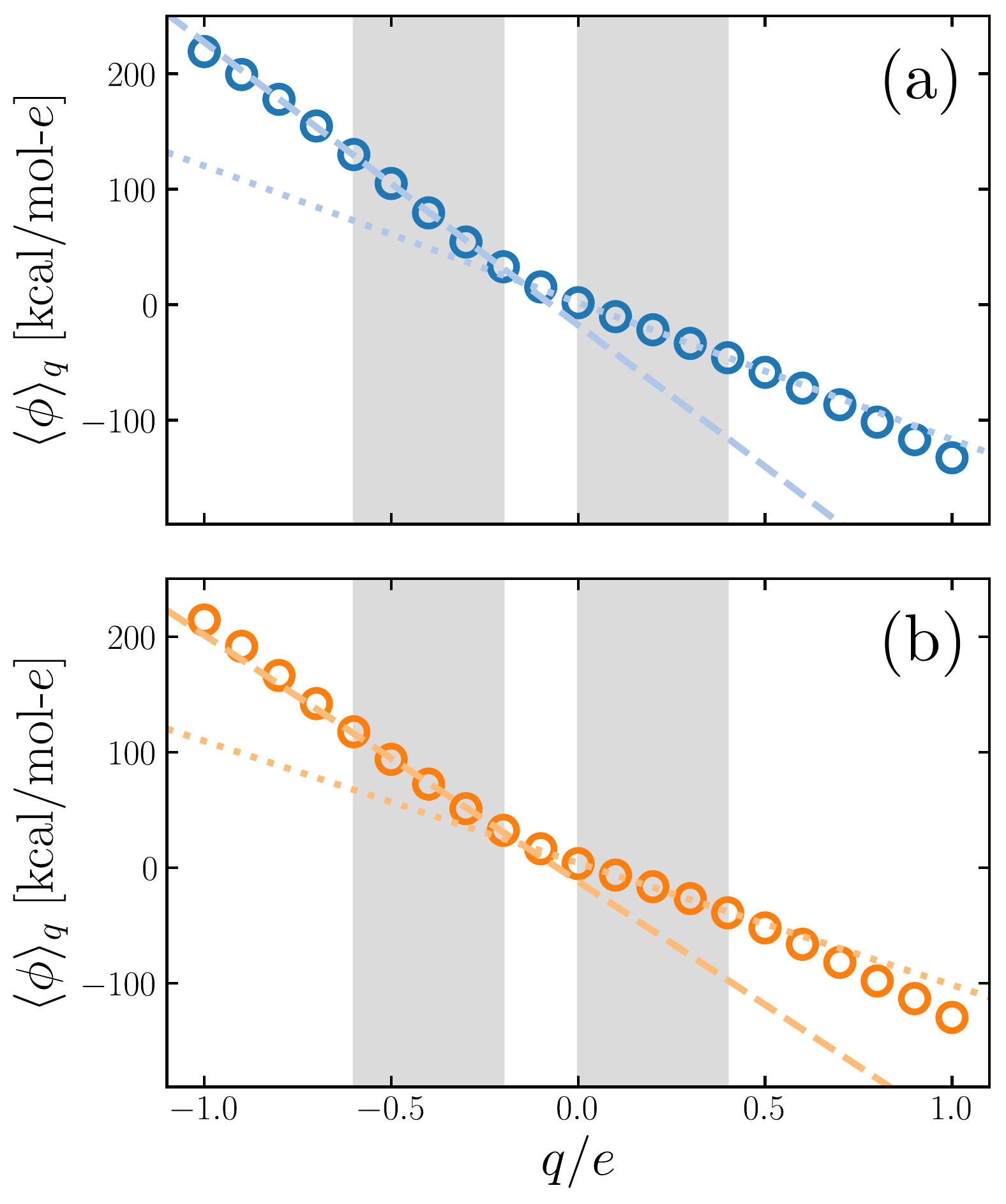}  
  \caption{$\langle\phi_{\rm solv}\rangle_q$ vs $q$ for $R=0.317$\,nm with the solute
    located (a) in bulk and (b) at the interface. The dashed and
    dotted lines show linear fits to the left and right shaded
    regions, respectively.}
  \label{fig:PLR3o17}
\end{figure}

\begin{figure}[tb]
  \includegraphics[width=7.5cm]{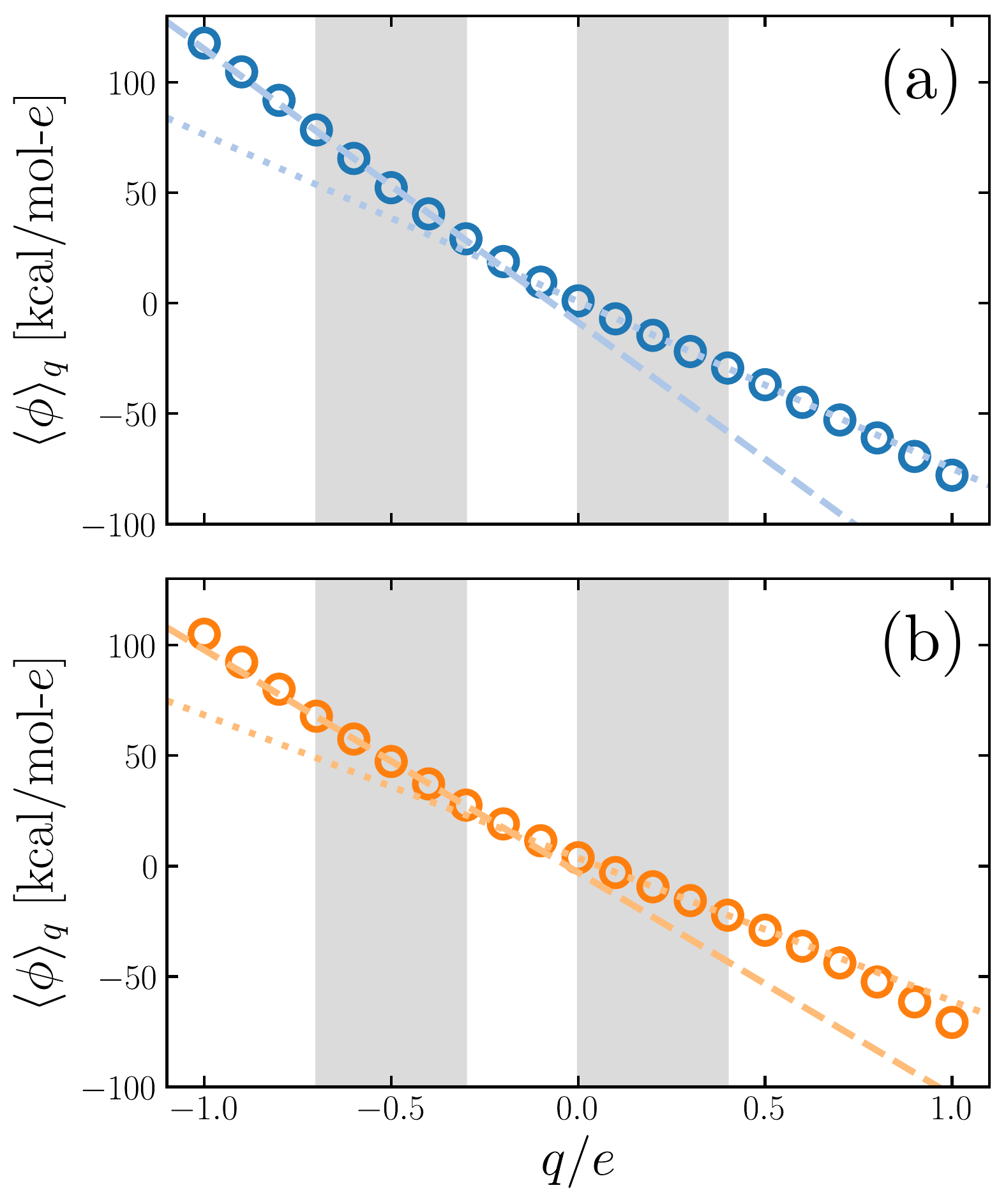}  
  \caption{$\langle\phi_{\rm solv}\rangle_q$ vs $q$ for $R=0.415$\,nm with the solute
    located (a) in bulk and (b) at the interface. The dashed and
    dotted lines show linear fits to the left and right shaded
    regions, respectively.}
  \label{fig:PLR4o145}
\end{figure}

Figures~\ref{fig:PLR2o4}, \ref{fig:PLR3o17} and~\ref{fig:PLR4o145}
show $\langle\phi_{\rm solv}\rangle_q$ vs $q$ for $R=0.240\,{\rm nm},
0.317\,{\rm nm}$ and~$0.415$\,nm, respectively, both for the solute in
bulk and at the interface. Note that these results have not been
corrected for the finite size of the simulation cell: we will correct
$\phi_{\rm neut}$ for finite size effects when computing $\Delta_{\rm
  ads}\psi^{\rm (PLR)}$, where other finite size effects largely
cancel \cite{cox2018interfacial}. For $R=0.317$\,nm and $R=0.415$\,nm
we can see that PLR is broadly reasonable for the solute in bulk, but
some small deviations are seen. These deviations are more pronounced
when the solute is at the interface. For $R=0.240$\,nm, the above PLR
model breaks down at large negative $q$, but it remains reasonable for
smaller values of the absolute charge. By fitting straight lines to
the anion and cation response, we can obtain values for $q_{\rm c}$,
$\langle(\delta\phi_{\rm solv})^2\rangle_+$ and
$\langle(\delta\phi_{\rm solv})^2\rangle_-$. The results from using
these in Eq.~\ref{eqn:PLR_psi} to compute $\Delta_{\rm ads}\psi^{\rm
  (PLR)}$ are presented in \tcr{Fig.~4b} in the main article. Results
for $R=0.75$\,nm and $R=1$\,nm are not shown because, while anion and
cation response do still differ, the degree of nonlinearity is much
less on an absolute scale than for the smaller solutes. This makes it
challenging to reliably obtain $q_{\rm c}$.

\clearpage

\section{Constructing $P_0(\phi_{\rm solv})$}

\begin{figure}[b]
  \includegraphics[width=7.5cm]{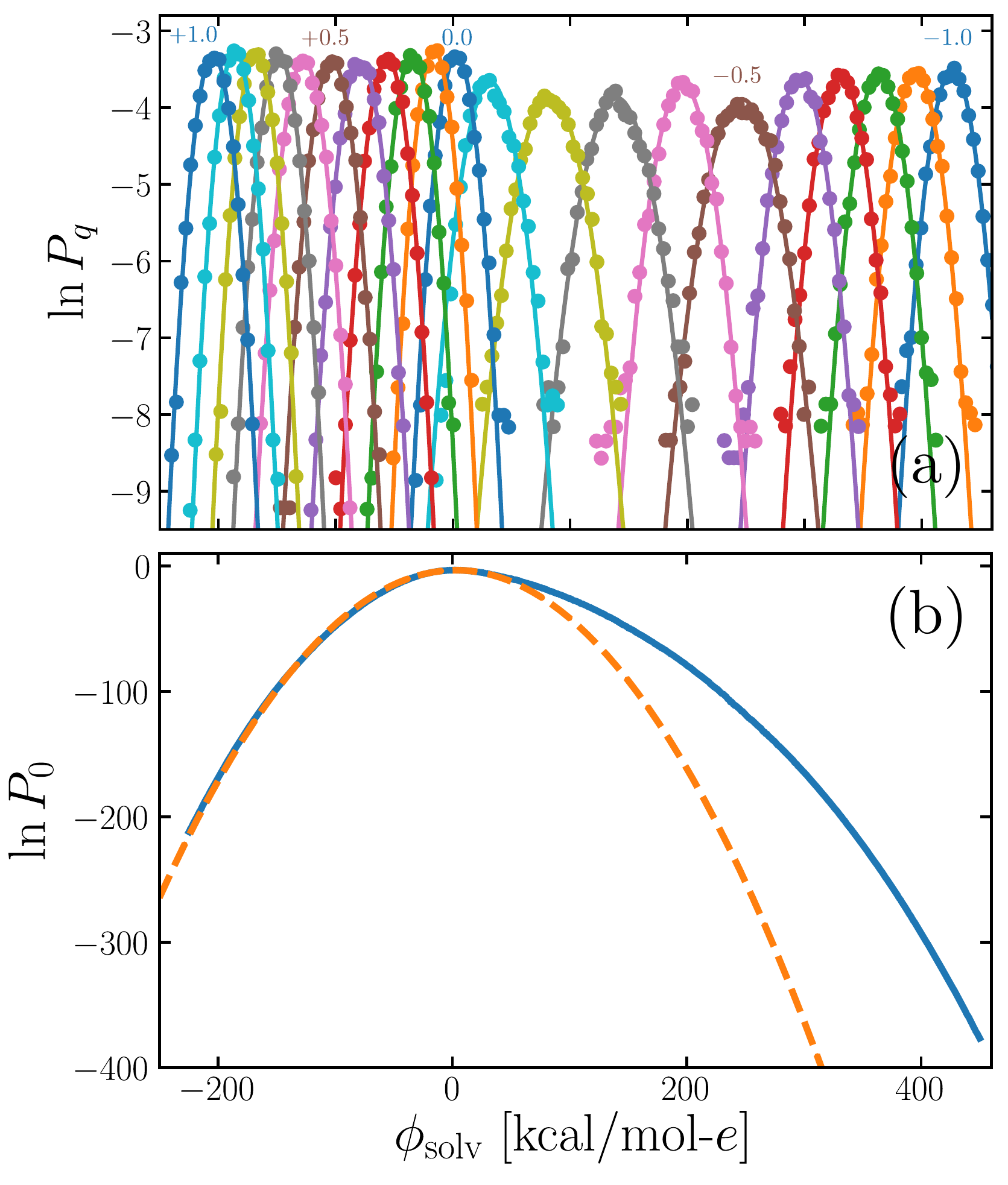}  
  \caption{(a) $P_q(\phi_{\rm solv})$ for
    $q/e=-1.0,-0.9,\ldots,0.0,\ldots,0.9,1.0$ with
    $R=0.240$\,nm. Solid lines indicate normalized Gaussian
    distributions and are included as a guide to the eye. (b)
    $P_0(\phi_{\rm solv})$ reconstructed from the set of $P_q$ using
    MBAR \cite{shirts2007accurate} (solid line). The dashed line
    indicates a normalized Gaussian distribution with mean and
    variance obtained from the simulation at $q/e=0$. Note that finite
    size corrections have not been applied to these plots.}
  \label{fig:Pq_composite}
\end{figure}

In order to compute $F_{\rm chg}(q)$ from \tcr{Eq.~7}, we require
$P_0(\phi_{\rm solv})$, the probability distribution of $\phi_{\rm
  solv}$. For the range of $q$ of interest, i.e. $-1 \le q/e \le 1$,
sampling $P_0$ directly (in the absence of solute charge) would yield
grossly insufficient data in the extreme wings of the
distribution. Instead, we obtain $P_0$ by histogram reweighting using
MBAR \cite{shirts2007accurate}. As an illustration,
Fig.~\ref{fig:Pq_composite}\,(a) shows probability distributions
$P_q(\phi_{\rm solv})$ of $\phi_{\rm solv}$ at the center of the
solute ($R=0.240$\,nm) with different values of $q$. Using data from
simulations across the full range of $q$, we then construct
$P_0(\phi_{\rm solv})$, as shown in Fig.~\ref{fig:Pq_composite}\,(b).

\end{document}